\documentclass[10pt,conference]{IEEEtran} 
\IEEEoverridecommandlockouts
\usepackage{cite}
\usepackage{amsmath,amssymb,amsfonts}
\usepackage{graphicx}
\usepackage{textcomp}
\usepackage{xcolor}
\usepackage{algorithm}
\usepackage{algorithmic}
\usepackage{amsmath}
\usepackage{multirow}
\usepackage{multicol}
\usepackage{subcaption}
\usepackage{booktabs}
\usepackage{xspace}
\usepackage{bm}
\usepackage{url}
\usepackage{enumitem}
\setlist[enumerate]{nolistsep}
\def\BibTeX{{\rm B\kern-.05em{\sc i\kern-.025em b}\kern-.08em
    T\kern-.1667em\lower.7ex\hbox{E}\kern-.125emX}}

\newcommand{\tool}{REPEAT\xspace}

\begin{document}

\title{Keeping Pace with Ever-Increasing Data: Towards Continual Learning of Code Intelligence Models}

\author{\IEEEauthorblockN{Shuzheng Gao$^{1}$, Hongyu Zhang$^{2}$, Cuiyun Gao$^{1\ast}$, Chaozheng Wang$^{1}$}

\IEEEauthorblockA{$^1$ School of Computer Science and Technology, Harbin Institute of Technology, Shenzhen, China}

\IEEEauthorblockA{$^2$ School of Big Data and Software Engineering, Chongqing University, China}



\IEEEauthorblockA{szgao98@gmail.com, hyzhang@cqu.edu.cn, gaocuiyun@hit.edu.cn, wangchaozheng@stu.hit.edu.cn}

\thanks{$^{\ast}$ Corresponding author. The author is also affiliated with Peng Cheng Laboratory and Guangdong Provincial Key Laboratory of Novel Security Intelligence Technologies.}

}

\pagestyle{plain}

\maketitle

\begin{abstract}
Previous research on code intelligence usually trains a deep learning model on a fixed dataset in an offline manner. 
However, in real-world scenarios, new code repositories emerge incessantly, and the carried new knowledge is beneficial for providing up-to-date code intelligence services to developers.
In this paper, 
we aim at the following problem: \textit{How to enable code intelligence models to continually learn from ever-increasing data?} 
One major challenge here 
is \textit{catastrophic forgetting}, meaning that the model can easily forget knowledge learned from previous datasets when learning from the new dataset. 
To tackle this challenge, we propose \tool, a novel method for continual learning of code intelligence models. Specifically, \tool addresses the catastrophic forgetting problem with representative exemplars replay and adaptive parameter regularization. The representative exemplars replay component selects informative and diverse exemplars in each dataset and uses them to retrain model periodically. The adaptive parameter regularization component recognizes important parameters in the model and adaptively penalizes their changes to preserve the knowledge learned before. We evaluate the proposed approach on three code intelligence tasks including code summarization, software vulnerability detection, and code clone detection. Extensive experiments demonstrate that \tool consistently outperforms baseline methods on all tasks. For example, \tool improves the conventional fine-tuning method by 1.22, 5.61, and 1.72 on code summarization, vulnerability detection and clone detection, respectively.
\end{abstract}


\section{Introduction}\label{sec:intro}
Recently, deep learning-based models have been widely utilized in many fields of software engineering, especially in the tasks associated with source code~\cite{DBLP:conf/iwpc/HuLXLJ18,DBLP:conf/sigsoft/WangYGP0L22,DBLP:conf/icse/GuZ018,DBLP:conf/ijcai/ZanCYLKGWCL22}. 
With the large-scale open source code corpora and the advanced deep learning techniques, these models achieve state-of-the-art performance on a variety of code intelligence tasks including code  summarization~\cite{DBLP:conf/iwpc/HuLXLJ18,DBLP:conf/acl/AhmadCRC20}, code clone detection~\cite{DBLP:journals/corr/abs-2002-08653,eva_clone}, and software vulnerability detection~\cite{DBLP:journals/tosem/ZouZXLJY21,DBLP:conf/nips/ZhouLSD019,DBLP:conf/ndss/LiZXO0WDZ18}, and thereby help improve the productivity of software developers. 


Despite the promising results, existing code-related tasks usually require to train a deep neural network model on a pre-collected and fixed dataset in an offline manner, which restricts their practicality in actual applications.
In real-world scenarios, the problem of code evolution exists in nature and the data are always produced in a continuous fashion, e.g., 71 million new repositories appear on GitHub from 2016 to 2018~\cite{timeline}. Besides, as indicated in \cite{Oraclejdk}, the number of JDK APIs increased from 211 to 4,403 in the past two decades. 
These new repositories and APIs may contain programming knowledge, which is crucial for providing up-to-date and high-quality assistance for developers.  
A neural network model that does not learn any knowledge from the new repositories cannot always give precise predictions on them~\cite{DBLP:conf/nips/BrownMRSKDNSSAA20,DBLP:journals/pami/LangeAMPJLST22}. 
A trained code intelligence model should be upgraded with the emergence of new datasets.
Therefore, enabling models with the ability to continuously learn knowledge over time is of vital importance for applying the code intelligence models in practice. As far as we know, little dedicated effort has been devoted to investigating this critical problem in code intelligence community.


\begin{figure}
    \centering
    \includegraphics[width=0.49\textwidth]{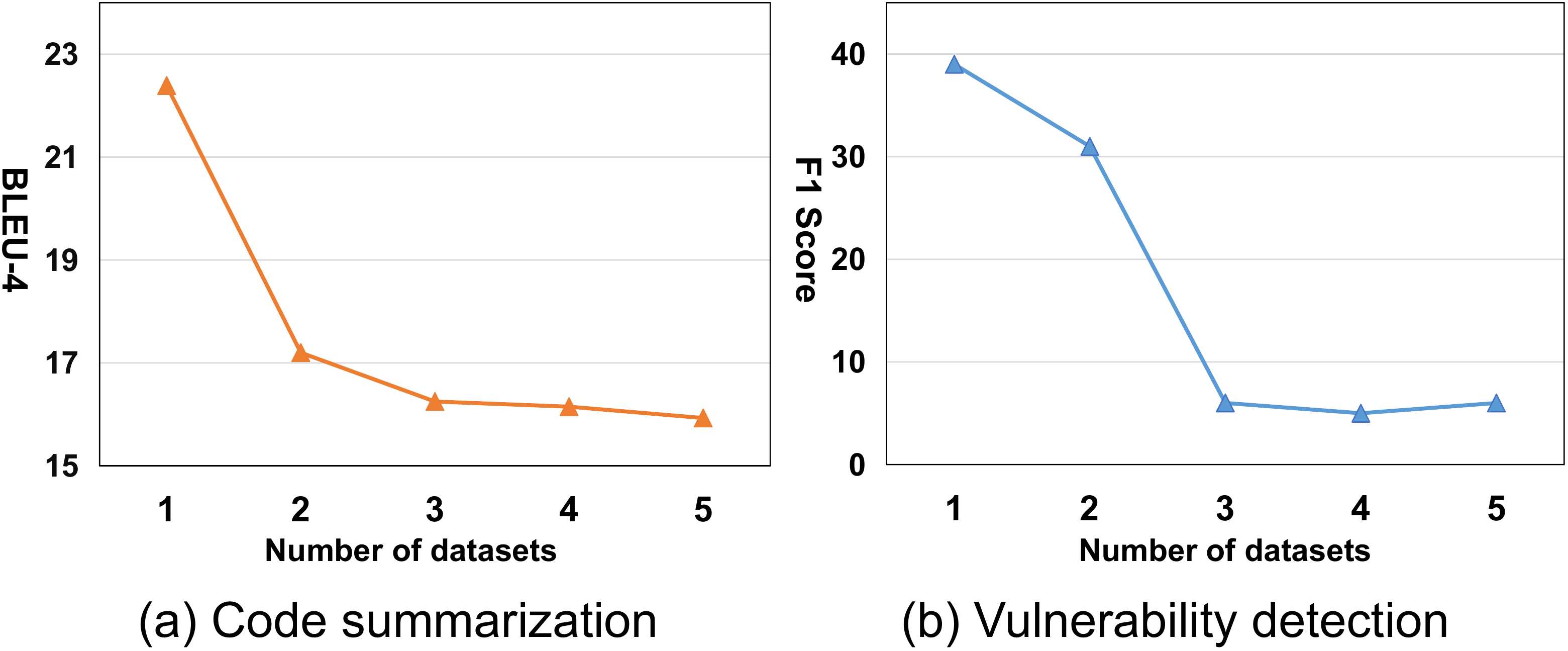}
    \caption{The performance degradation on the first dataset after fine-tuning CodeBERT on new datasets. The datasets of code summarization and vulnerability detection are from CodeSearchNet and Big-Vul respectively. For more details refer to Section~\ref{subsec:dataset}.}
    \label{fig:example}
\end{figure}


One straightforward solution to maintain code intelligence models is to
fine-tune the models on new datasets~\cite{DBLP:conf/emnlp/FengGTDFGS0LJZ20,DBLP:conf/emnlp/0034WJH21}. However, this
method suffers from the \textit{catastrophic forgetting} problem, where the model forgets the knowledge learned from previous data. As shown in Figure~\ref{fig:example} (b), 
directly fine-tuning on new datasets (i.e., datasets 2-5) leads to obvious performance degradation on the first dataset, e.g., showing a 28.9\% and 84.6\% drop after training on the fifth dataset
for code summarization and vulnerability detection, respectively.
Another straightforward solution
is to retrain the model on all historical datasets when a new dataset is available. Although the method can preserve the model performance,
it is often infeasible in practice due to the severe computation overhead~\cite{DBLP:journals/pami/LangeAMPJLST22,DBLP:journals/pami/LiH18a}. For example, the training time of the state-of-the-art code intelligence model CodeT5 on the dataset with 8 million instances is 12 days~\cite{DBLP:conf/emnlp/0034WJH21}.
Considering the generally larger data in practice, timely model update is difficult.
Therefore, it is challenging to mitigate the catastrophic forgetting problem while avoiding enormous training costs.

\begin{figure}[t]
     \centering
    \includegraphics[width=0.48 \textwidth]{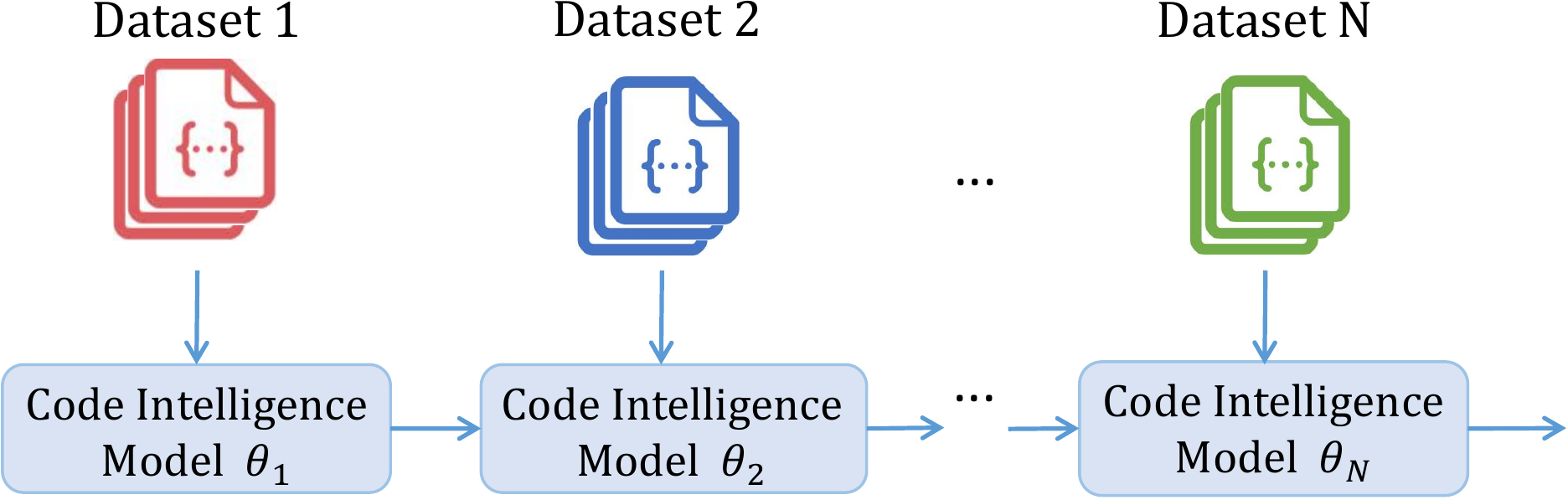}
    \caption{Continual learning paradigm.}
    \label{fig:overview}
\end{figure}

To mitigate the challenge, we propose to resort to the \textit{continual learning} techniques~\cite{DBLP:journals/pami/LangeAMPJLST22,chen2018lifelong} in the machine learning field. Continual learning, as shown in Figure~\ref{fig:overview}, is capable of learning from a sequence of newly-added datasets, while alleviating the catastrophic forgetting problem with limited extra cost.
Recently, a variety of techniques for continual learning have been proposed~\cite{DBLP:conf/naacl/WangXYGCW19,DBLP:conf/acl/HanDGLLLSZ20,DBLP:conf/emnlp/MiCZHF20,DBLP:conf/www/YuanYHCWC22}. 
Among them, replay-based and regularization-based methods have drawn substantial attention, where the former retrains the models with a fixed set of exemplars selected from previously-seen datasets and the latter regularizes the changes of models' parameters.
For example, EMR~\cite{DBLP:conf/naacl/WangXYGCW19} is a typical replay-based method that randomly selects exemplars from previous datasets. EWC~\cite{kirkpatrick2017overcoming} is a popular regularization-based method that employs elastic weight consolidation to measure the importance of each parameter through the Fisher information. 
However, EMR and EWC do not take the characteristics of data (such as the existence of patterns 
and noisy data)
into account, while these data characteristics are common in code intelligence tasks~\cite{DBLP:conf/oopsla/Allamanis19,DBLP:conf/icse/SunLL0022,DBLP:journals/corr/abs-2207-05579}. 



In this paper, we present \tool, a method with \textbf{R}epresentative \textbf{E}xemplars replay and adaptive \textbf{P}aram\textbf{E}ter regulariz\textbf{AT}ion, for enabling continual learning in code intelligence models.
Specifically, due to the existence of various data patterns and noisy data in code intelligence datasets~\cite{DBLP:conf/icse/SunLL0022}, instead of randomly selecting exemplars like what EMR does, we propose to preserve the most representative (i.e., informative and diverse) samples in each dataset for model retraining.
Besides, considering the popularity of
code reuse in software development~\cite{DBLP:journals/tse/KamiyaKI02,DBLP:conf/sigsoft/KimSN05},
we propose 
an adaptive parameter regularization mechanism
to control the degree of parameter updating based on the sharing knowledge between datasets.

We conduct experiments on three popular code comprehension tasks including code summarization, software vulnerability detection, and code clone detection. For simulating the continual learning scenarios, we conduct data partitioning according to the project information and produce a sequence of datasets.
Experimental results on two state-of-the-art code intelligence models CodeBERT and CodeT5 demonstrate that \tool can achieve superior performance in continual code intelligence scenarios, and effectively mitigate catastrophic forgetting problems.

The contributions of this work can be summarized as:
\begin{enumerate}
    \item To the best of our knowledge, we are the first to explore the performance of code intelligence models under the continual learning scenarios.
    \item We propose \tool, a novel continual learning-based method with representative exemplars replay and adaptive parameter regularization, to prevent code intelligence models from catastrophically forgetting the learned knowledge.
    \item We conduct extensive experiments on three code intelligence tasks with two state-of-the-art 
    models. Experimental results demonstrate the effectiveness of \tool and its ability to mitigate the catastrophic forgetting problem.
\end{enumerate}

\section{Background and Related Work}\label{sec:back}

\subsection{Code Intelligence}
Deep learning techniques have been widely used in many code intelligence tasks. 
In this section, we introduce the code intelligence tasks covered in our work, including code summarization, vulnerability detection, and code clone detection. In addition, we also introduce the background of pre-trained models of code.

\textbf{Code Summarization}
aims to automatically generate a short natural language description that can accurately summarize the functionality of the given code snippet~\cite{DBLP:conf/acl/IyerKCZ16,DBLP:conf/iwpc/HuLXLJ18}. 
Recent studies~\cite{DBLP:conf/acl/AhmadCRC20,DBLP:conf/acl/WuZZ21} in this field resort to deep learning techniques and formulate the code summarization task as a sequence-to-sequence neural machine translation (NMT) problem. By adopting the advanced NMT frameworks and integrating the source code properties, these models~\cite{DBLP:conf/acl/WuZZ21,DBLP:conf/icse/TangSLGHZ022,10.1145/3522674} achieve state-of-the-art performance on this task.

\textbf{Vulnerability Detection} is the task of identifying whether the given source code contains  vulnerabilities such as resource leaks~\cite{DBLP:journals/corr/abs-2102-04664,DBLP:journals/corr/abs-2212-14274}. It is crucial for the safety of a software system~\cite{DBLP:journals/tosem/ZouZXLJY21,DBLP:conf/nips/ZhouLSD019}. In general, this task is formulated as a binary classification task. Specifically, the model first encodes the code snippet into representation vector. Then a classifier is employed to predict the probability of vulnerability. The model is usually trained by minimizing the cross-entropy loss function.

\textbf{Code Clone Detection}
is the task of measuring the similarity between two code snippets which can help reduce the cost of software development and  maintenance~\cite{DBLP:journals/corr/abs-2002-08653,eva_clone}. It contains two sub-tasks including binary classification and code-to-code retrieval~\cite{DBLP:journals/corr/abs-2102-04664}. In this work, we follow the mainstream of studies~\cite{DBLP:conf/ijcai/WeiL17,DBLP:conf/iwpc/YuLCLXW19,DBLP:journals/corr/abs-2002-08653} and focus on the former sub-task that detects whether two given code snippets have the same functionality.

\textbf{Pre-trained models of code.}
Recently, a series of work~\cite{DBLP:conf/emnlp/FengGTDFGS0LJZ20,DBLP:conf/iclr/GuoRLFT0ZDSFTDC21,DBLP:conf/emnlp/0034WJH21} leverages the self-supervised pre-training techniques and substantially improves the performance in many downstream tasks. The improvement attributes to unleashing the power of large-scale unlabelled code corpora. For example, CodeBERT~\cite{DBLP:conf/emnlp/FengGTDFGS0LJZ20} is a pioneer work in code intelligence based on masked language modeling and replaced token detection tasks. The more recent work CodeT5~\cite{DBLP:conf/emnlp/0034WJH21} formulates all the tasks in a sequence to sequence paradigm with different task-specific prefixes and achieves state-of-the-art performance on a variety of code intelligence tasks.

\subsection{Continual Learning}
Continual learning, also known as incremental learning or lifelong learning, aims at learning from sequential data stream without catastrophic forgetting~\cite{mccloskey1989catastrophic}. Existing work in continual learning can be mainly divided into three categories: replay-based methods~\cite{DBLP:conf/nips/Lopez-PazR17,DBLP:conf/iclr/SunHL20}, regularization-based methods~\cite{kirkpatrick2017overcoming,DBLP:journals/pami/LiH18a} and architecture-based methods~\cite{DBLP:conf/cvpr/AljundiCT17,DBLP:conf/cvpr/MallyaL18}. Among them, architecture-based methods dynamically allocate new components for new datasets, which increases the model's parameters dramatically when the number of datasets becomes extremely large. This property contradicts with our scenario where enormous datasets may be involved. Therefore, in this work, we only focus on the research in the first two categories.

\textbf{Replay-based methods} mitigate catastrophic forgetting by storing a limited set of exemplars from previous datasets and using them to retrain the model periodically. It has been proven to be the most promising method in NLP~\cite{DBLP:conf/acl/HanDGLLLSZ20,DBLP:conf/naacl/WangXYGCW19}. EMR~\cite{DBLP:conf/naacl/WangXYGCW19} is a typical replay-based method that randomly selects saved exemplars in each dataset. It mixes saved exemplars from previous datasets and the new dataset to fine-tune the model. Therefore, instead of random selection, how to select exemplars with higher quality become a great challenge. 
Recently, a variety of works utilize the characteristic of their own tasks to guide the exemplars selection process. For example, ARPER~\cite{DBLP:conf/emnlp/MiCZHF20} selects representative exemplars based on the number of slots for the dialog system. Ramalho and Garnelo~\cite{DBLP:conf/iclr/RamalhoG19} propose to store exemplars that the model has less confidence in. EMAR~\cite{DBLP:conf/acl/HanDGLLLSZ20} selects diverse exemplars that can cover different relation patterns for relation extraction. Other methods~\cite{DBLP:conf/iclr/SunHL20,DBLP:conf/iclr/KemkerK18} mainly focus on the storage cost and employ a generator that can generate pseudo examples for each task. Different from these works, we focus on how to leverage the characteristics of code intelligence tasks and select high-quality exemplars for them.

\textbf{Regularization-based methods} preserve learned knowledge from old datasets by constraining the model's parameters from changing too much from the previous model. Compared with replay-based methods, these methods are more efficient and can provide an effective supplement for them since the constraint of replay-based methods might be not strong enough due to the limited exemplars size~\cite{DBLP:conf/emnlp/MiCZHF20,DBLP:journals/pami/LangeAMPJLST22}. 
For example, EWC~\cite{kirkpatrick2017overcoming}, as the most representative regularization-based method, employs elastic weight consolidation to regularize the changes in the parameter, where the importance of each parameter is measured by the Fisher information. It has been widely used in many works~\cite{DBLP:conf/emnlp/LiQH21,DBLP:conf/www/YuanYHCWC22,DBLP:conf/issta/YuanZHFHHY22} in NLP. Recently, Mi et al.~\cite{DBLP:conf/emnlp/MiCZHF20} and Yuan et al.~\cite{DBLP:conf/www/YuanYHCWC22} further propose to utilize the vocabulary size and question syntax information to improve EWC for dialogue and question generation respectively. In our work, we aim to better adopt EWC in code intelligence tasks by improving its flexibility based on the common code reuse in the software development process.

\section{Methodology}\label{sec:approach}
\subsection{Problem Formulation}
Following previous continual learning research~\cite{DBLP:conf/www/YuanYHCWC22,DBLP:conf/emnlp/MiCZHF20}, we formulate the continual learning setting of code intelligence tasks as follows. In the continual learning setting, the models are required to be
trained
on a sequence of datasets. 
We assume that there are $N$ datasets in total, where each dataset consists of a training set $T_i$, a validation set $V_i$, and a test set $S_i$, i.e., $D_{1:N} = \{\{T_1, V_1, S_1\}, \{T_2, V_2, S_2\},..., \{T_N, V_N, S_N\}\}$. At the $i$-{th} time step, the model will be trained and validated on $T_i$ and $V_i$, respectively. After training on the $i$-{th} dataset,
the model is expected to
still perform well on both the $i$-th dataset and all the previous datasets. Therefore, 
the model evaluation is performed
on the test sets of all learned dataset, e.g., $\{S_1, S_2,..., S_i\}$. Besides, as adopted in many previous work~\cite{DBLP:conf/emnlp/MiCZHF20,DBLP:conf/emnlp/LiQH21,DBLP:conf/acl/HanDGLLLSZ20}, we further assume that when the model is trained on the $i$-th dataset, it cannot directly access the data from previous datasets. But a pre-stored exemplar set $E_{1:i-1}$ with limited and fixed size is permitted, which is also practical in real scenario since using all previous data will bring extensive training and storage cost.

\subsection{Proposed Approach}
\begin{figure}
    \centering
    \includegraphics[width=0.45\textwidth]{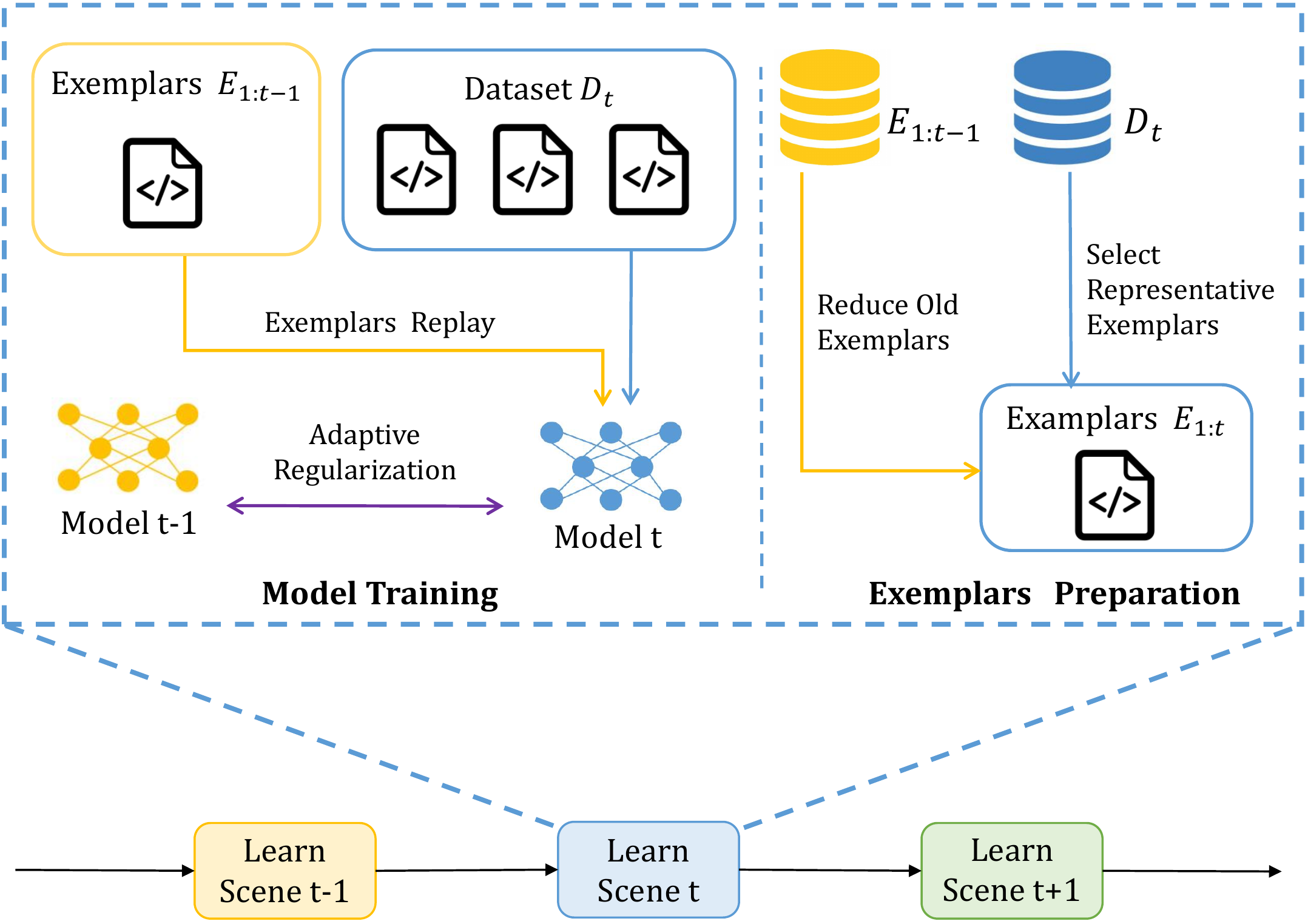}
    \caption{The overview of \tool.}
    \label{fig:approach}
\end{figure}

In this section, we introduce how we mitigate the catastrophically forgetting problem in continual code intelligence tasks. As shown in Figure~\ref{fig:approach}, in the training stage, apart from fine-tuning on current dataset, our method contains two extra components, including \textit{representative exemplars replay} which retrains model with representative samples of previous datasets and \textit{adaptive parameter regularization} which further regularizes the model's parameter change adaptively. After training, \tool selects representative samples in current dataset and involves them to construct new exemplars.

\subsubsection{Representative Exemplars Reply}
To prevent the models from catastrophically forgetting the knowledge in previous datasets, one popular and effective method is to retrain the model on a limited subset of previous datasets~\cite{DBLP:conf/nips/Lopez-PazR17,DBLP:conf/iclr/SunHL20}. Specifically, when training the model on the dataset $D_t$, we combine $D_t$ with the saved exemplars from all previous datasets, denoted as $E_{1:t-1}=\{E_1, E_2,..., E_{t-1}\}$, and use them together to retrain the model. Formally, the training object for dataset $t$ can be formulated as follows:
\begin{equation}
L^{replay}_t = \sum_{\{x,y\} \in D_t \cup E_{1:t-1}}^{} L(x, y, \theta_t)
\end{equation}
where $L$ is the loss function for each task such as cross-entropy and $\theta_t$ is the parameters of the model trained on the $t$-th dataset. The size of saved exemplars should be as small as possible to reduce the computation, i.e., $E_{1:t-1} \ll D_{1:t-1}$.

The quality of the saved exemplars is vital to the effectiveness of the replay-based method. Therefore, instead of selecting samples at random as EMR~\cite{DBLP:conf/naacl/WangXYGCW19}, we propose to preserve the most representative samples in each dataset to replay. Specifically, considering the existence of various code patterns and noisy data in code intelligence datasets, we hold the view that the representative exemplars are both diverse and informative. 
Specifically, as shown in Algorithm~\ref{alg:framework}, after the training stage, we select the exemplars as follows. 

\textbf{Diverse Exemplars Selection:} 
Data in code intelligence tasks often share similarities, e.g., code snippets share similar functionalities or code patterns~\cite{DBLP:conf/naacl/LeClairM19,DBLP:conf/oopsla/Allamanis19}. Thus, our exemplars should be diverse and cover various kinds of data instances. Meanwhile, the redundancy among the exemplars should be reduced. To this end, we propose to employ a $K$-means-based algorithm to divide  the vector representation of  samples into $K$ clusters (Lines 1-6), and then select samples from each cluster. In our experiments, we utilize TF-IDF (Term Frequency - Inverse Document Frequency) for the vectorization, which has been shown efficient and effective in text and code retrieval~\cite{DBLP:conf/kbse/WeiLLXJ20,ramos2003using}.

\textbf{Informative Exemplars Selection:} Previous work has demonstrated that noisy samples exist in many popular benchmarks such as CodeSearchNet~\cite{DBLP:conf/icse/SunLL0022,DBLP:journals/corr/abs-2207-05579}. To enable the model to learn valuable knowledge from the exemplars, we select the informative samples rather than the noisy samples for replay. 
According to recent work~\cite{DBLP:conf/nips/HanYYNXHTS18,DBLP:conf/iccv/HuangQJZ19}, the noisy data are generally with higher training loss. We propose to select informative exemplars by filtering the high-loss samples. Specifically, as shown in Lines 7-12 in Algorithm~\ref{alg:framework}, in each cluster, we first filter the possible noisy data by selecting $\mu \cdot m_i$ samples with the lowest loss values as our candidate exemplars, and then randomly select $m_i$ samples from them. Here $\mu$ is a hyper-parameter to control the size of candidate exemplars and $m_i$ denotes the number of samples to be preserved in the $i$-th cluster. 

\textbf{Previous Exemplars Removal:} To maintain a fixed size of exemplars, 
after selecting new exemplars in the current dataset, we should also reduce the preserved exemplars from previous datasets. Specifically, assume that the total size of replayed exemplars is set to $M$ and we are working on the $t$-th dataset. We propose to keep the exemplars from each dataset with the same size. Specifically, for each previous dataset, we remove $\frac{M}{t-1}-\frac{M}{t}$ (the rounding operation is omitted for simplicity) samples with the highest loss calculated before (Lines 14-16). In this way, we maintain a fixed size of exemplars that contain the most representative exemplars.
\begin{algorithm}[t]
\caption{Algorithm of exemplar selection}
\label{alg:framework}
\begin{algorithmic}[1]
\REQUIRE Current dataset $C$, previous exemplars $E_{1:t-1}$, number of task $t$, size of exemplars $M$, hyper-parameter $\mu$, clutser number $K$, model trained after the $t$-th dataset $\theta_t$\\
\ENSURE Newly saved exemplars $E_{1:t}$\\
\STATE Vectorize all samples in $C$ 
\STATE Apply $K$-means algorithm to divide samples in $C$ into $K$ clusters
\STATE $E_t\gets\varnothing$ 
\FOR{each cluster $c_i$ in $C$}
\STATE $m_i$ = $\frac{M}{t} \cdot \frac{\left | c_i \right | }{\left | C \right | }$ 
\STATE $P\gets\varnothing$ 
\FOR{each sample $\{x_{ij}, y_{ij}\}$ in $c_i$}
\STATE $P.\textit{insert}(\{x_{ij}, y_{ij}, L(x_{ij}, y_{ij}, \theta_t)\})$ 
\ENDFOR
\STATE Sort samples in $P$ based on $L(x_{ij}, y_{ij}, \theta_t)$ in ascending order
\STATE $P_i\gets$ \ Randomly select $m_i$ samples from $P_{1:\mu \cdot m_i}$
\STATE $E_t\gets E_t \cup P_i$ 
\ENDFOR
\FOR{each dataset $i$ in exemplars $E_{1:t-1}$}
\STATE Sort samples in $E_i$ based on $L$ in descending order
\STATE Remove first $\frac{M}{t-1}-\frac{M}{t}$ samples in $E_i$
\ENDFOR
\STATE $E_{1:t}\gets E_{1:t-1} \cup E_t$
\RETURN Newly saved exemplars $E_{1:t}$

\end{algorithmic}
\end{algorithm}


\subsubsection{Adaptive Parameter Regularization}
Although the \textit{Representative Exemplars Reply} component is able to alleviate the catastrophically forgetting problem by replaying representative samples, the model may overfit on the
saved samples and  forget other samples due to the limited exemplar size~\cite{DBLP:conf/acl/HanDGLLLSZ20}. 
Therefore, apart from exemplars replaying, we further propose to adapt
regularization on parameter change based on the Elastic Weight Consolidation (EMC) method~\cite{kirkpatrick2017overcoming}.

EWC measures the importance of each parameter to previous datasets and penalizes the parameter change by adding an elastic $L_2$ regularization term.
Specifically, the loss function of involving the EWC regularization for dataset $t$ is formulated as follows:

\begin{equation}
L_t = L^{replay}_t+\lambda\sum_{i}^{P}F_i \cdot (\theta_{t,i}-\theta_{t-1,i})^2
\end{equation}
where $P$ is the number of training parameters and $\lambda$ is a hyperparameter to balance the degree of preserving and learning,  
e.g., the model trained with larger $\lambda$ tends to preserve previous knowledge rather than learn new knowledge. $F_i=\nabla^2 L(x,y|\theta_{t-1,i})$ measures the importance of $i$-th parameter in model $\theta_{t-1}$ through the Fisher information matrix~\cite{kirkpatrick2017overcoming}, where $\nabla^2$ is the Laplace operator. 

Considering the popularity of software reuse in the process of software development~\cite{DBLP:journals/tse/KamiyaKI02,DBLP:conf/sigsoft/KimSN05}, there exist many similar code snippets in different datasets that share similar knowledge.
Intuitively, if the current dataset $D_t$ is similar with the previous datasets $D_{1:t-1}$, the model will only require sightly parameter updating since most knowledge in $D_t$ has been learned before.
The vanilla EWC method employs the same $\lambda$ for different datasets. However, this is not flexible for code intelligence tasks since the more current dataset is similar with previous ones, the stronger regularization is supposed to be applied. 
To this end, we propose an adaptive parameter regularization method to adjust the regularization adaptively. Specifically,  
we aggregate the TF-IDF vectors of all samples in the dataset to vectorize the whole dataset. With the vectors, we then obtain the adaptive regularization term with cosine similarity:

\begin{equation}\label{equ:adaptive}
\lambda = {\lambda}_{base} \cdot Cosine(V(D_t), V(E_{1:t-1}))
\end{equation}
where $V(\cdot)$ denotes the TF-IDF vectorization function. When calculating the adaptive regularization term and the value of $F_i$, we only utilize the saved exemplars rather than using all historical data to prevent extensive training costs.
\section{Experimental setup}\label{sec:setup}

\begin{figure}
    \centering
    \includegraphics[width=0.48\textwidth]{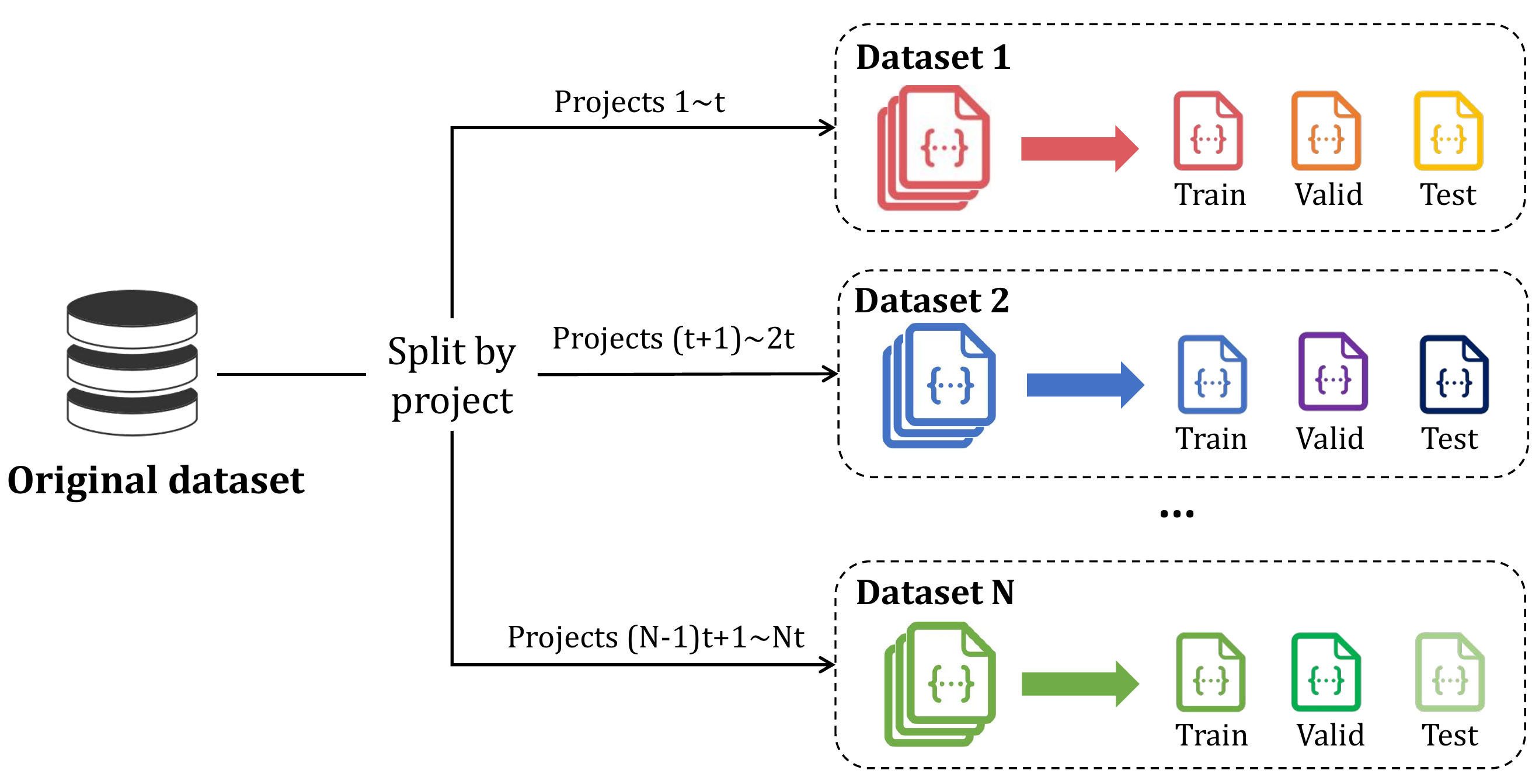}
    \caption{Illustration of dataset construction process. There is no overlapping between datasets.}
    \label{fig:dataset}
\end{figure}

\begin{table*}[t]
    \centering
    \caption{Statistics of the datasets used in this paper. CS, VD and CD denote code summarization, vulnerability detection and clone detection respectively.}
 \scalebox{0.96}{
    \begin{tabular}{cc|rrr|rrr|rrr|rrr|rrr}
    \toprule
    \multicolumn{2}{c|}{{\multirow{2}{*}{Task}}} & \multicolumn{3}{c|}{1st} & \multicolumn{3}{c|}{2nd} & \multicolumn{3}{c|}{3rd} & \multicolumn{3}{c|}{4th} & \multicolumn{3}{c}{5th}\\
     & & Train & Valid & Test & Train & Valid & Test & Train & Valid & Test & Train & Valid & Test & Train & Valid & Test\\
    \midrule
    \multirow{6}{*}{CS} & Java & 27,519 & 3,439 & 3,441 & 28,137 & 3,517 & 3,518 & 39,450 & 3,681 & 3,682 & 31,887 & 3,985 & 3,987 & 27,854 & 3,481 & 3,483\\
    & Python & 44,782 & 5,597 & 5,599 & 43,691 & 5,461 & 5,462 & 47,000 & 5,875 & 5,876 & 44,128 & 5,516 & 5,516 & 44,919 & 5,614 & 5,616\\
    & Go & 29,992 & 3,749 & 3,749 & 29,283 & 3,660 & 3,661 & 31,823 & 3,977 & 3,979 & 26,665 & 3,333 & 3,334 & 28,424 & 3,553 & 3,553\\
    & PHP & 44,341 & 5,542 & 5,544 & 42,351 & 5,293 & 5,295 & 41,405 & 5,175 & 5,177 & 40,433 & 5,054 & 5,055 & 46,057 & 5,757 & 5,758\\
    & Javascript & 9,962 & 1,245 & 1,246 & 10,748 & 1,343 & 1,345 & 10,541 & 1,317 & 1,319 & 10,185 & 1,273 & 1,274 & 10,722 & 1,340 & 1,341\\
    & Ruby & 4,544 & 568 & 568 & 4,398 & 549 & 551 & 4,259 & 532 & 533 & 4,520 & 565 & 566 & 4,348 & 543 & 544\\
    \midrule
    \multicolumn{2}{c|}{VD} & 66,551 & 8,318 & 8,320 & 9,465 & 1,183 & 1,184 & 50,818 & 6,352 & 6,353 & 8,324 & 1,040 & 1,042 & 15,748 & 1,968 & 1,970\\
    \midrule
    \multicolumn{2}{c|}{CD} & 40,000 & 5,000 & 5,000 & 40,000 & 5,000 & 5,000 & 40,000 & 5,000 & 5,000 & 40,000 & 5,000 & 5,000 & 40,000 & 5,000 & 5,000\\
    \bottomrule
    \end{tabular}
    \label{tab:dataset}}
 \end{table*}
 
\subsection{Datasets}\label{subsec:dataset}
So far, there has been no dataset for evaluating code intelligence tasks under the continual learning setting. Besides, the time information is missing in most code intelligence datasets. Therefore, to simulate the data-increasing process, we build upon existing datasets that involve the project information and construct a project-level continual learning version for them. 

\subsubsection{Code Summarization}
We employ the popular CodeSearchNet (CSN)~\cite{DBLP:journals/corr/abs-1909-09436} dataset which contains thousands of code snippets in six programming languages. 
We build a project-level continual learning version denoted as CSN-PC with the following steps. As shown in Figure~\ref{fig:dataset}, we first merge the data from the original dataset and extract the project of each method. Then, we randomly split the projects into five parts with the same size to simulate the data stream. For each part, we randomly split them into training set, validation set, and test set in a proportion of 8:1:1. Different datasets contain code snippet from different projects and there is no overlapping between datasets.
 
\subsubsection{Vulnerability Detection}
The dataset that we build upon is provided by Fan et al.~\cite{fan}, namely Big-Vul. It consists C/C++ code snippets in over 300 GitHub projects from 2002 to 2019 in Common Vulnerabilities and Exposures (CVE) database. We also construct a project-level continual learning version for this dataset, which is denoted as Big-Vul-PC. The construction procedure is the same as CSN-PC. For data in each part, we also randomly partition it into training, validation, and test sets with the ratio of 8:1:1.
\subsubsection{Clone Detection}
For code clone detection, we use the POJ dataset~\cite{DBLP:conf/aaai/MouLZWJ16} which contains 52,000 code snippets of C language with 104 functionalities as the evaluation dataset. Similarly, we also divide the 104 functionalities into 5 parts. 
Due to enormous data, e.g., the size of training data in one part is $\frac{8,400 \times 8,400}{2}=35,280,000$, we follow the strategy in \cite{eva_clone,DBLP:conf/ijcai/WeiL17} and randomly sample 40,000/5,000/5,000 code snippets for the training set, validation set, and test set respectively.

We present the detailed statistics of each processed dataset in Table~\ref{tab:dataset}.

\subsection{Evaluation Metrics}

For code summarization, we follow previous work~\cite{DBLP:conf/acl/AhmadCRC20,DBLP:conf/iclr/LiuCXS021,DBLP:conf/icse/ZhangW00020} and use three popular metrics BLEU-4~\cite{DBLP:conf/acl/PapineniRWZ02}, ROUGE-L~\cite{lin-2004-rouge} and METEOR~\cite{DBLP:conf/acl/BanerjeeL05} for evaluation. 

BLEU measures the similarity of two summaries by calculating the ratio of $N$ groups of word similarity between them. A higher BLEU score indicates higher similarity. We follow previous work~\cite{DBLP:conf/acl/AhmadCRC20,DBLP:conf/icse/ZhangW00020} and use BLEU-4 for evaluation. It is computed as:
\begin{equation}
     BLEU-4 = BP \times \exp(\sum_{n=1}^{4}\tau_n \log P_n),
\end{equation}
where $P_n$ is the ratio of $n$-gram in the prediction summary that are also in the reference summary. $BP$ is the brevity penalty and $\tau_n$ is set to $1/4$.

METEOR evaluates generated summaries by aligning them to the reference summaries and calculating the similarity scores as follows:
\begin{equation}
    METEOR = (1-\gamma \cdot \textit{frag}^\beta) \cdot \frac{P \cdot R}{\alpha \cdot P + (1-\alpha) \cdot R},
\end{equation}
where P and R are unigram precision and recall, \textit{frag} is the fragmentation fraction. $\alpha$, $\beta$ and $\gamma$ are three penalty parameters whose default values are 0.9, 3.0, and 0.5, respectively.

ROUGE-L calculates the F-score based on Longest Common Subsequence (LCS) between two summaries. Given a generated summary $X$ and the reference summary $Y$, ROUGE-L is computed as:
\begin{equation}
     P_{lcs} = \frac{LCS(X,Y)}{n}, \quad  R_{lcs} = \frac{LCS(X,Y)}{m},
\end{equation}
\begin{equation}
     F_{lcs} = \frac{(1+\beta ^2)P_{lcs}R_{lcs}}{R_{lcs}+\beta^2 P_{lcs}},
\end{equation}
where $m$ and $n$ are the length of $X$ and $Y$, respectively. $\beta = P_{lcs}/R_{lcs}$ and $F_{lcs}$ is the computed ROUGE-L score.




\subsubsection{Vulnerability  Detection and Clone Detection}
For vulnerability detection and clone detection, we follow previous work~\cite{DBLP:conf/nips/ZhouLSD019,DBLP:conf/emnlp/0034WJH21} and evaluate the results by Precision~(P), Recall~(R), and F1:

\begin{equation}
P = \frac{TP}{TP+FP},R = \frac{TP}{TP+FN},F1 =  \frac{2 \cdot P \cdot  R}{P+R}
\end{equation}
where TP, FP, TN, and FN denote the number of true positives, false positives, true negatives, and false negatives respectively. Since the datasets of vulnerability detection and clone detection are highly imbalanced, e,g., the ratio of vulnerable and non-vulnerable code in Big-Vul is about 1:16, the results of the F1 score are more preferable for these two tasks.

\subsubsection{Evaluation for Continual Learning}
To better evaluate the effectiveness of each method under the continual learning setting, following~\cite{DBLP:conf/emnlp/MiCZHF20,DBLP:conf/emnlp/LiQH21}, we further evaluate them by their average performance on all test sets:

\begin{equation}\label{equ:ave}
\Omega = \frac{1}{K}\sum_{i=1}^{K} \Omega_{i}, \quad  \Omega_{i} = \frac{1}{i}\sum_{j=1}^{i} \Omega_{j,i}
\end{equation}
where $\Omega_{j,i}$ denotes the performance on the $j$-th test set after the $i$-{th} dataset has been learned. Here $\Omega$ can represent any metric we introduced above. $\Omega$ evaluates the model’s overall performance on all historical datasets. A method with a higher value of $\Omega$ should perform well on both the current dataset and all previous datasets.


\subsection{Baselines}
We evaluate the performance of each continual learning method with two state-of-the-art pre-trained models of code, CodeBERT~\cite{DBLP:conf/emnlp/FengGTDFGS0LJZ20} and CodeT5~\cite{DBLP:conf/emnlp/0034WJH21}. CodeBERT is an encoder-only pre-trained model that achieves promising results on code intelligence tasks. 
CodeT5 is an encoder-decoder model that formulates all the tasks in a sequence to sequence paradigm with different prefixes. It involves two code-related pre-training objectives, i.e., identifier tagging and masked identifier prediction, and achieves state-of-the-art performance on a variety of code intelligence tasks.

For the continual learning methods, we follow previous work~\cite{DBLP:conf/emnlp/MiCZHF20,DBLP:conf/www/YuanYHCWC22} and compare \tool with the following methods.  \textbf{FT} is a straightforward method that directly fine-tunes the model on each new task. It always serves as a lower bound in the field of continual learning.  \textbf{EMR}~\cite{DBLP:conf/naacl/WangXYGCW19} is a typical replay-based method that simply retrains the model with old samples randomly selected from previous tasks. Instead of replaying past samples, \textbf{EWC}~\cite{kirkpatrick2017overcoming} employs elastic weight consolidation to regularize the changes in parameters, where the importance of each parameter is measured by the Fisher information. \textbf{Upper} train the model with data from the current dataset and all historical datasets which can provide the upper bound for our evaluation.

\begin{table*}[t]
    \centering
    \caption{Experimental results of code summarization. BLEU and ROUGE denote BLEU-4 and ROUGE-L respectively.
    }
 \scalebox{0.88}{
    \begin{tabular}{cc|cccccccccccc|cc}
    \toprule
    \multicolumn{2}{c|}{{\multirow{2}{*}{\textbf{Approach}}}} & \multicolumn{2}{c}{Java} & \multicolumn{2}{c}{Python} & \multicolumn{2}{c}{Go} & \multicolumn{2}{c}{PHP} & \multicolumn{2}{c}{Javascript} & \multicolumn{2}{c|}{Ruby}  & \multicolumn{2}{c}{Average}\\
    \cmidrule{3-16} 
    &  & BLEU & ROUGE & BLEU & ROUGE & BLEU & ROUGE & BLEU & ROUGE & BLEU & ROUGE & BLEU & ROUGE & BLEU & ROUGE\\
    \midrule
    \multirow{5}{*}{CodeBERT} & \multicolumn{1}{|c|}{Upper} & 23.43 & 39.67 & 21.18 & 35.73 & 35.74 & 51.88 & 27.00 & 40.21 & 16.85 & 24.06 & 17.52 & 26.43 & 23.62 & 36.33\\
    \cmidrule{2-16} 
    \multicolumn{1}{c|}{} & FT & 20.32 & 35.83 & 18.28 & 31.08 & 29.64 & 46.43 & 24.46 & 36.77 & 15.96 & 22.28 & 16.34 & 25.17 & 20.83 & 32.93\\
    \multicolumn{1}{c|}{} & \tool & \textbf{21.47} & \textbf{37.16} & \textbf{19.38} & \textbf{32.95} & \textbf{33.11} & \textbf{49.15} & \textbf{25.13} & \textbf{37.48} & \textbf{16.26} & \textbf{22.79} & \textbf{16.95} & \textbf{26.12} & \textbf{22.05} & \textbf{34.28}\\
    \cmidrule{2-16} 
    \multicolumn{1}{c|}{} & EMR & 21.06 & 36.59 & 18.99 & 32.15 & 32.45 & 48.64 & 24.71 & 37.24 & 16.12 & 22.60 & 16.65 & 25.81 & 21.66 & 33.83\\
    \multicolumn{1}{c|}{} & EWC & 20.37 & 36.47 & 18.81 & 32.75 & 30.64 & 47.52 & 24.74 & 37.46 & 16.00 & 22.58 & 16.32 & 25.32 & 21.15 & 33.68\\
    \midrule
    \multirow{5}{*}{CodeT5} & \multicolumn{1}{|c|}{Upper} & 26.66 & 43.58 & 23.28 & 39.23 & 38.55 & 54.81 & 28.81 & 43.49 & 18.69 & 29.75 & 20.30 & 32.98 & 26.05 & 40.64\\
    \cmidrule{2-16} 
    \multicolumn{1}{c|}{} & FT  & 23.58 & 40.41 & 21.26 & 36.74 & 34.59 & 51.49 & 26.29 & 40.63 & 17.43 & 28.67 & 18.94 & 31.40 & 23.68 & 38.22\\
    \multicolumn{1}{c|}{} & \tool & \textbf{24.79} & \textbf{41.44} & \textbf{21.99} & \textbf{37.59} & \textbf{36.10} & \textbf{52.55} & \textbf{26.84} & \textbf{41.12} & \textbf{17.96} & \textbf{28.73} & \textbf{19.71} & \textbf{32.10} & \textbf{24.57} & \textbf{38.92} \\
    \cmidrule{2-16} 
    \multicolumn{1}{c|}{} & EMR & 24.12 & 40.89 & 21.52 & 36.94 & 35.76 & 52.40 & 26.55 & 40.99 & 17.65 & 28.39 & 19.15 & 31.75 & 24.13 & 38.56\\
    \multicolumn{1}{c|}{} & EWC & 23.89 & 40.87 & 21.49 & 37.20 & 34.81 & 51.43 & 26.58 & 41.07 & 17.43 & 28.15 & 19.03 & 31.41 & 23.87 & 38.36\\
    \bottomrule
    \end{tabular}
    \label{tab:sum}}
\end{table*}

\subsection{Implementation Details}
We reproduce the results of CodeBERT and CodeT5 based on the official repository released by the authors. We train the models with the default hyperparameters in CodeBERT and CodeT5 such as learning rate and optimizer. As for the training epochs, to ensure coverage, we set them to 15, 10, and 5 for code summarization, vulnerability detection, and clone detection, respectively. We use the same hyper-parameters for all continual learning methods for fair comparison. When applying our exemplar selection methods to the classification task, we conduct algorithm~\ref{alg:framework} for each class respectively.

In experiments, we set the size of replayed exemplars to 1\% of the whole training data for comparison, which is considerably smaller than the whole size of training data. For the parameter $\lambda$, we set it as 2000 by default. Both hyper-parameters $K$ and $\mu$ in the exemplars preparation process are set to 5. We discuss the impact of the above parameters in Section~\ref{sec:RQ4}. 

All the experiments are conducted on a server (Ubuntu 20.04) with 4 Nvidia Tesla V100 GPUs which have 32 GB graphic memory.

\section{Experimental Results}\label{sec:result}

\begin{table}[t]
    \centering
    \caption{Experimental results of vulnerability  detection. }
 \scalebox{1}{
    \begin{tabular}{c|ccc|ccc}
    \toprule
    \multirow{2}{*}{\textbf{Approach}} & \multicolumn{3}{c|}{CodeBERT} & \multicolumn{3}{c}{CodeT5} \\
    \cmidrule{2-7}
    & F1 & P & R & F1 & P & R \\
    \midrule
    Upper & 36.98 & 43.29 & 32.12 & 80.18 & 76.03 & 85.05  \\
    \midrule
    FT & 26.86 & 41.91 & 21.98 & 78.48 & 70.58 & \textbf{88.63} \\
    \tool & \textbf{32.47} & \textbf{44.61} & \textbf{27.12} & \textbf{79.49} & \textbf{73.00} & 87.58   \\
    \midrule
    EMR & 28.42 & 42.58 & 22.62 & 78.79 & 72.43 & 86.89   \\
    EWC & 27.47 & 42.34 & 21.35 & 78.51 & 72.92 & 85.56  \\
    \bottomrule
    \end{tabular}
    \label{tab:defect}}
\end{table}

\begin{table}[t]
    \centering
    \caption{Experimental results of clone detection. }
 \scalebox{1.05}{
    \begin{tabular}{c|ccc|ccc}
    \toprule
    \multirow{2}{*}{\textbf{Approach}} & \multicolumn{3}{c|}{CodeBERT} & \multicolumn{3}{c}{CodeT5} \\
    \cmidrule{2-7}
     & F1 & P & R & F1 & P & R \\
    \midrule
    Upper & 92.10 & 95.78 & 88.57 & 95.04 & 92.96 & 97.24 \\
    \midrule
    FT  & 86.89 & 86.87 & \textbf{87.27} & 85.72 & 91.50 & 82.03\\
    \tool  & \textbf{88.61} & \textbf{91.07} & 86.37 & \textbf{91.26} & 91.05 & \textbf{91.77} \\
    \midrule
    EMR & 88.52 & 90.53 & 86.83 & 90.44 & 90.95 & 90.26 \\
    EWC & 86.53 & 89.82 & 83.94 & 86.56 & \textbf{91.83} & 82.96\\
    \bottomrule
    \end{tabular}
    \label{tab:clone}}
\end{table}

In this section, we conduct experiments to evaluate the performance of \tool. We mainly focus on the following research questions:

\begin{enumerate}[label=\bfseries RQ\arabic*:,leftmargin=.5in]
    \item How effective is \tool compared to baseline methods?
    \item Does \tool improve model's generalization to unseen projects?
    \item What is the impact of each component on the performance of \tool?
    \item How does \tool perform under different parameter settings?
\end{enumerate}

\begin{table*}[t]
    \centering
    \caption{Experimental results of performance on unseen dataset. }
 \scalebox{1}{
    \begin{tabular}{c|ccccccccc}
    \toprule
    \multirow{2}{*}{\textbf{Approach}} & \multicolumn{3}{c}{Code Summarization} & \multicolumn{3}{c}{Vulnerability detection} & \multicolumn{3}{c}{Clone detection} \\
    \cmidrule{2-10} 
    & BLEU-4 & METEOR & ROUGE-L & F1 &  Precision & Recall & F1 & Precision & Recall  \\
    \midrule
    FT  & 14.73  &  11.34 & 28.14 & 8.22 & 33.33 & 4.69 & 80.40 & 77.73 & 83.26\\
    EMR  & 14.48  &  11.28 & 28.56 & 8.14 & 31.03 & 4.69 & 84.80 & 86.84 & 82.85\\
    EWC  & 14.64 & 10.91 & 28.69 & 6.64 & \textbf{36.84} & 3.65 & 84.94 & \textbf{91.75} & 79.08\\
    \midrule
    \tool-1  & 14.47  &  10.57 & 28.14 & 10.40 & 22.41 & 6.77 & 74.25 & 70.99 & 77.82\\
    \tool-2  & 14.64  &  10.92 & 28.55 & 7.11 & 14.75 & 4.69 & 81.47 & 79.37 & 83.68\\
    \tool-3  & 14.56  &  11.12 & 29.03 & 7.86 & 24.32 & 4.69 & 85.53 & 88.39 & \textbf{82.85}\\
    \midrule
    \tool  & \textbf{14.97}  & \textbf{11.55} & \textbf{29.37} & \textbf{10.79} & 17.44 & \textbf{7.51} & \textbf{85.90} & 89.19 & \textbf{82.85}\\
    \bottomrule
    \end{tabular}
    \label{tab:generalization}}
\end{table*}

\begin{table*}[t]
    \centering
    \caption{Ablation study. \tool-C, \tool-L and \tool-A denote removing clustering-based exemplars selection, loss-based exemplars selection and adaptive regularization respectively. BLEU and ROUGE denote BLEU-4 and ROUGE-L respectively.}
 \scalebox{0.95}{
    \begin{tabular}{cc|cccccccccccc}
    \toprule
    \multicolumn{2}{c|}{{\multirow{2}{*}{\textbf{Approach}}}} & \multicolumn{3}{c}{Code Summarization (Java)} & \multicolumn{3}{c}{Code Summarization (Python)} & \multicolumn{3}{c}{Vulnerability Detection} & \multicolumn{3}{c}{Clone Detection} \\
    \cmidrule{3-14} 
     &  &BLEU & METEOR & ROUGE & BLEU & METEOR & ROUGE &  F1 & Precision & Recall & F1 & Precision & Recall \\
    \midrule
    \multirow{5}{*}{CodeBERT} & \multicolumn{1}{|c|}{\tool}  & \textbf{21.47}  & \textbf{14.93} & \textbf{37.16} &  \textbf{19.38}  & 13.02 & \textbf{32.95} & \textbf{32.47} & 44.61 & \textbf{27.12} & \textbf{88.61} & \textbf{91.07} & 86.37\\
    \cmidrule{2-14}
    \multicolumn{1}{c|}{}& \multicolumn{1}{l|}{\tool-C} & 21.34 & 14.86 & 37.03 & 18.99 & 12.86 & 32.75 & 30.48 & 43.24 & 24.96 & 88.34 & 89.55 & 87.74\\
    \multicolumn{1}{c|}{}& \multicolumn{1}{l|}{\tool-L} & 20.83 & 14.54 & 36.68 & 18.88 & 12.84 & 32.02 & 32.35 & \textbf{45.45} & 26.56 & 88.49 & 90.86  & \textbf{86.38}\\
    \multicolumn{1}{c|}{}& \multicolumn{1}{l|}{\tool-A} & 21.30 & 14.84 & 37.05 & 19.29 & \textbf{13.06} & 32.50 & 31.88 & 44.25 & 25.81 & 88.27 & 90.43 & 85.61\\
    \midrule
    \multirow{5}{*}{CodeT5} & \multicolumn{1}{|c|}{\tool}  & \textbf{24.79}  & 18.52 & 41.44 &  \textbf{21.99}  & \textbf{16.73} & \textbf{37.59} & \textbf{79.49} & 73.00 & \textbf{87.58} & \textbf{91.26} & \textbf{91.05} & \textbf{91.77}\\
    \cmidrule{2-14}
    \multicolumn{1}{c|}{}& \multicolumn{1}{l|}{\tool-C} & 24.64 & \textbf{18.55} & 41.37 & 21.39 & 15.98 & 36.92 & 79.22 & 73.29 & 86.54 & 89.04 & 90.33 & 88.14\\
    \multicolumn{1}{c|}{}& \multicolumn{1}{l|}{\tool-L} & 24.31 & 18.36 & 41.13 & 21.27 & 16.38 & 36.71 & 78.87 & \textbf{74.74} & 84.01 & 90.16 & 91.99 & \textbf{88.91}\\
    \multicolumn{1}{c|}{}& \multicolumn{1}{l|}{\tool-A} & 24.69 & 18.54 & \textbf{41.46} & 21.35 & 15.86 & 37.12 & 79.32 & 73.60 & 86.30 & 89.44 & 91.46 & 87.94\\
    \bottomrule
    \end{tabular}
    \label{tab:ablation}}
\end{table*}


\subsection{RQ1: Comparison with Baselines}\label{sec:RQ1}
\begin{figure}
    \centering
    \includegraphics[width=0.5\textwidth]{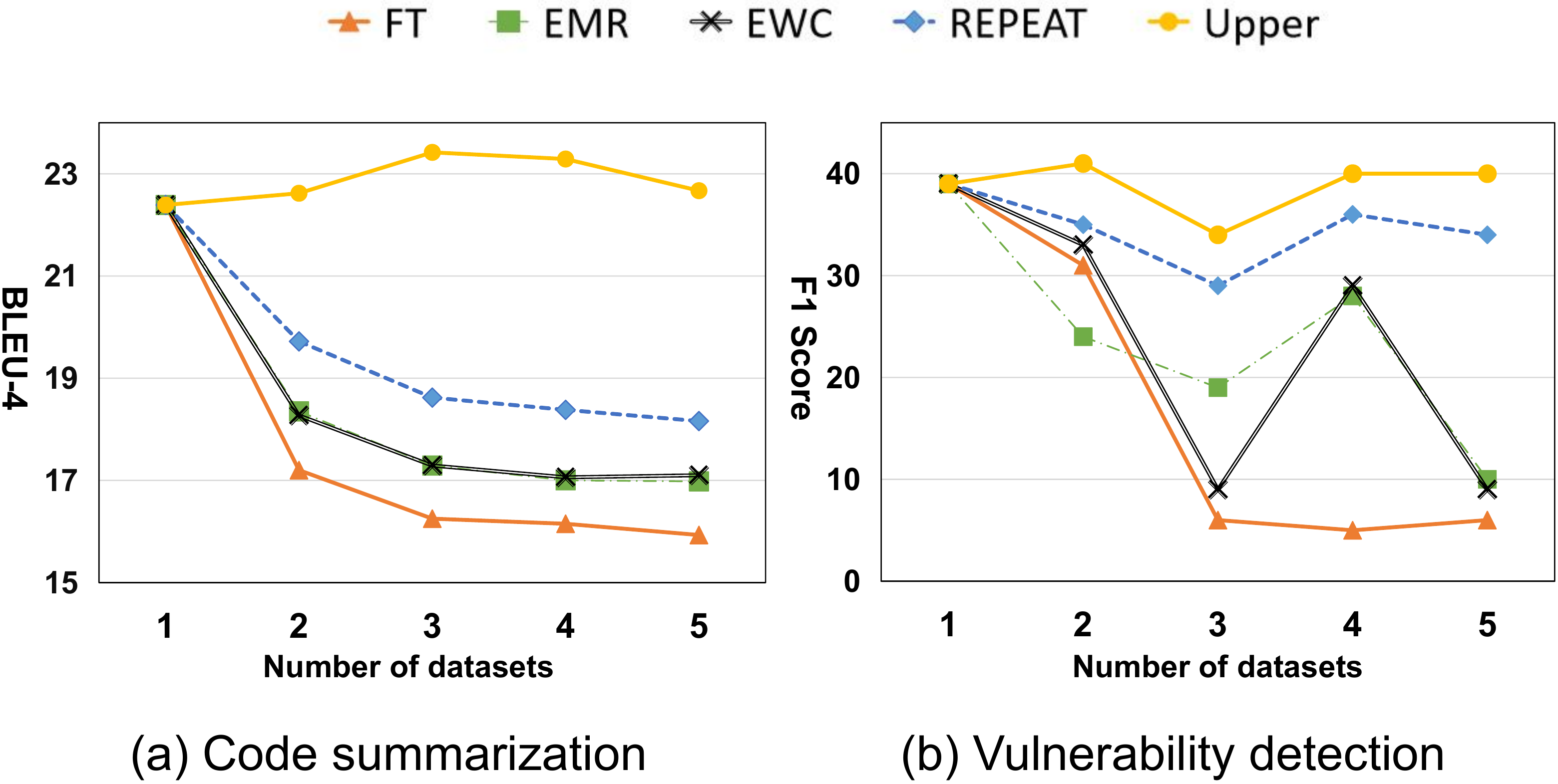}
    \caption{Performance of each method on the first dataset of (a) code summarization and (b) vulnerability detection.}
    \label{fig:first}
\end{figure}

We evaluate the performance of \tool with three code intelligence tasks and present the experimental results in Tables~\ref{tab:sum}-\ref{tab:clone}. Due to the page limit, we only present the final results calculated by Equ~\ref{equ:ave}. And for the code summarization task, we only present the BLEU-4 and ROUGE-L in Table~\ref{tab:sum}. The results for the METEOR metrics and detailed results on each dataset can be found in the GitHub repository\footnote{{\url{https://github.com/ReliableCoding/REPEAT}}}. We further show the trends of code summarization and vulnerability detection in Figure~\ref{fig:first}. Based on these results, we summarize the following findings:

\textbf{FT severely suffers from catastrophic forgetting problem.} As can be seen in the trends in Figure~\ref{fig:first}, with the arrival of new datasets, the performance of FT on the first dataset drops obviously. Specifically, after training on the fifth dataset, its performance on the first dataset drops 7.46 and 33.28 on code summarization and vulnerability detection in terms of BLEU-4 and F1, respectively.

\textbf{Exemplars replay and parameter regularization are beneficial for alleviating forgetting.} By comparing EMR and EWC with FT, we can find that exemplars replay and parameter regularization bring an obvious improvement on all tasks. Specifically, on clone detection, EMR and EWC outperform FT by 4.72 and 0.84 regarding the F1 metrics on CodeT5, respectively. The trends in Figure~\ref{fig:first} also indicate that EMR and EWC can effectively alleviate the catastrophic forgetting problem.

\textbf{The proposed \tool is effective in different continual code intelligence tasks.} As shown in Table~\ref{tab:sum}-\ref{tab:clone},  \tool can consistently achieve the best performance on all metrics and tasks. For example, on code summarization, \tool improves FT by 1.22 and 0.89 points respectively regarding the average BLEU-4 and ROUGE metrics on CodeBERT. When compared with EMR and EWC, \tool also dramatically improves them by at least 4.05 and 0.70 points regarding the F1 score on vulnerability detection for CodeBERT and CodeT5 respectively.
As shown in Figure~\ref{fig:first}, the trends also indicate that \tool can better mitigate the catastrophic forgetting problem and achieve the best performance among all methods. We also notice that there is an unnatural jump on the F1 score for the fourth dataset of vulnerability detection. We suppose it might be caused by the data distribution of different datasets and will explore this phenomenon in depth in the future. 

\subsection{RQ2: Generalization Evaluation}\label{sec:RQ2}
In this section, we conduct experiments to study whether learning knowledge continually can improve the model's generalization on unseen projects. To investigate this problem, we train all baselines and \tool on the first four datasets and evaluate their performance on the fifth dataset. Apart from them, we further involve \tool-1, \tool-2, and \tool-3 which represent training \tool on the first, second, and third dataset, respectively. We use CodeBERT as our base model and select Java as the evaluation dataset for code summarization. From the results in Table~\ref{tab:generalization}, we have the following observations:

\textbf{Continual learning is beneficial to the generalization on unseen projects.} Comparing \tool with \tool-1, \tool-2, and \tool-3, we can find that \tool consistently outperforms them on all tasks. Specifically, on code summarization, \tool improves them by at least 0.33, 0.43, and 0.34 points in terms of the BLEU-4, METEOR, and ROUGE-L metrics, respectively. This indicates that continually training the model with new data can benefit the model's generalization on unseen projects.

\textbf{\tool can achieve better generalization performance.} As shown in Table~\ref{tab:generalization}, compared with FT, EMR, and EWC, \tool achieves the best result on all tasks. Specifically, on vulnerability detection and clone detection, \tool improves the best baseline by 2.65 and 0.96 points regarding the F1 score, which indicates that \tool can achieve better generalization performance by better preserving and reusing knowledge learned before.

\begin{figure*}[t]
    \centering
      \begin{subfigure}[b]{0.24\textwidth}
        \centering
        \includegraphics[width=1\textwidth]{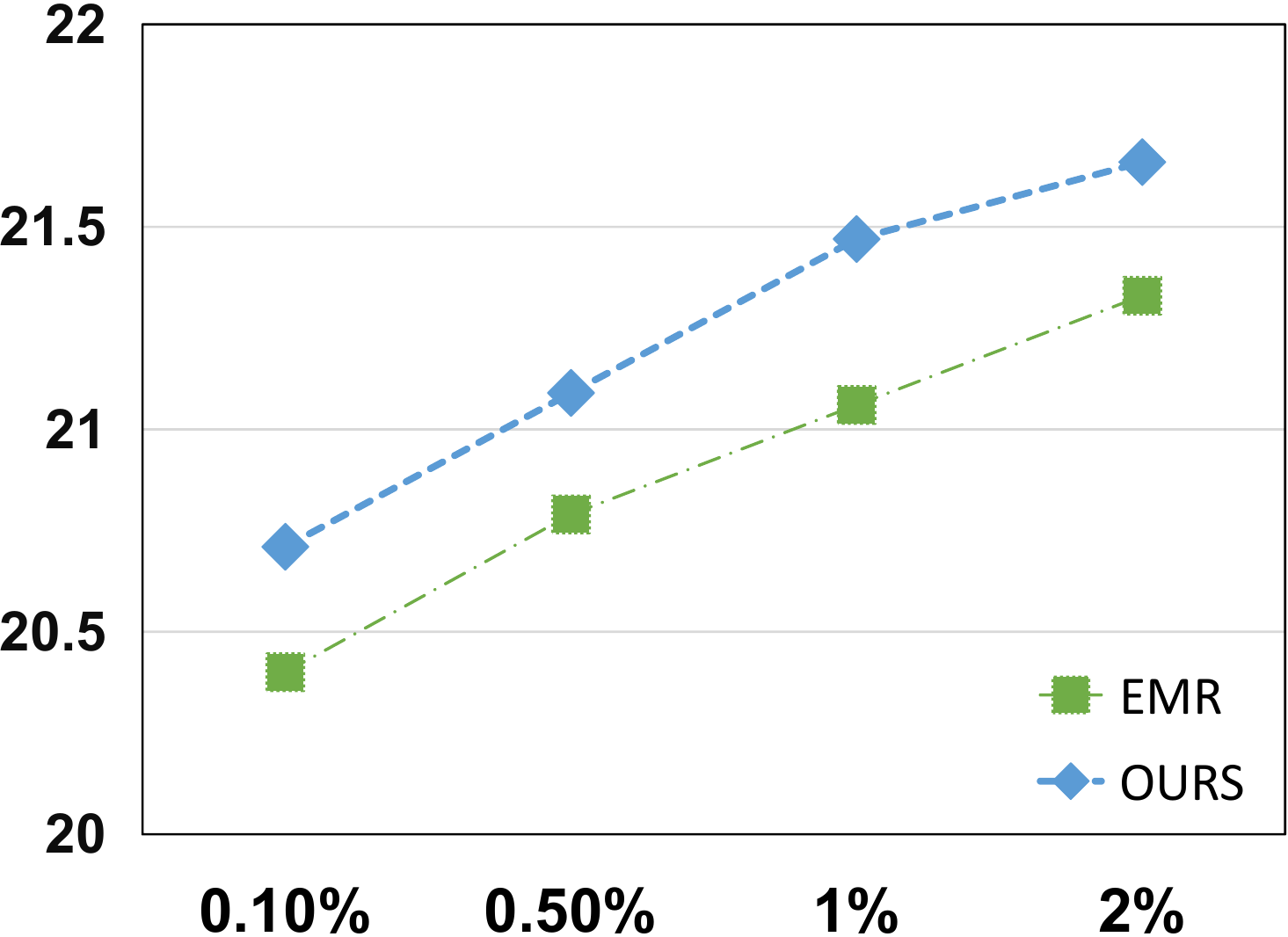}
        \caption{Exemplars size of CodeBERT.}
        \label{tab:size-cb}
      \end{subfigure}
      \hfill
      \begin{subfigure}[b]{0.24\textwidth}
        \centering
        \includegraphics[width=1\textwidth]{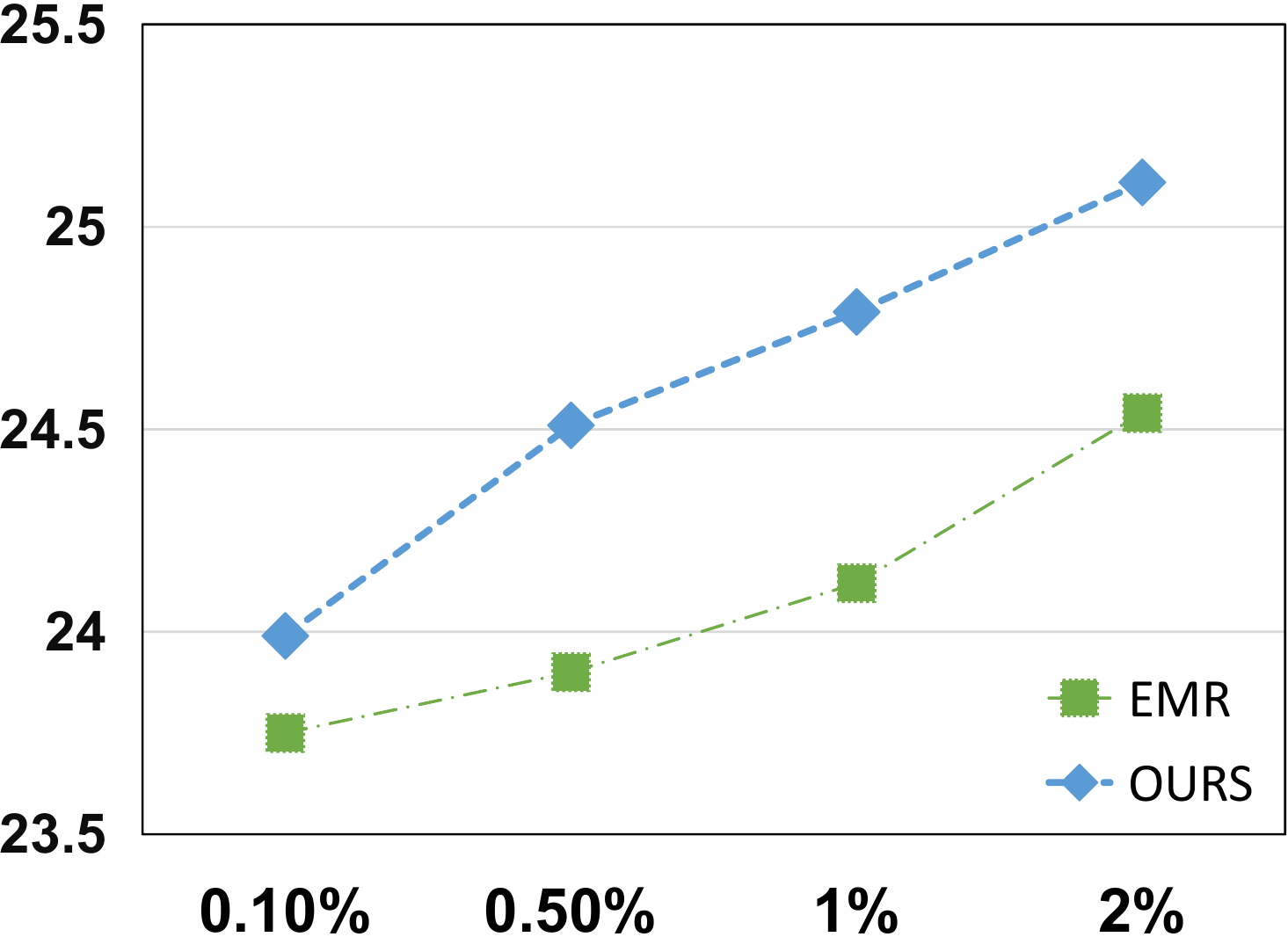}
        \caption{Exemplars size of CodeT5.}
        \label{tab:size-ct}
      \end{subfigure}
      \hfill
      \begin{subfigure}[b]{0.24\textwidth}
        \centering
        \includegraphics[width=1\textwidth]{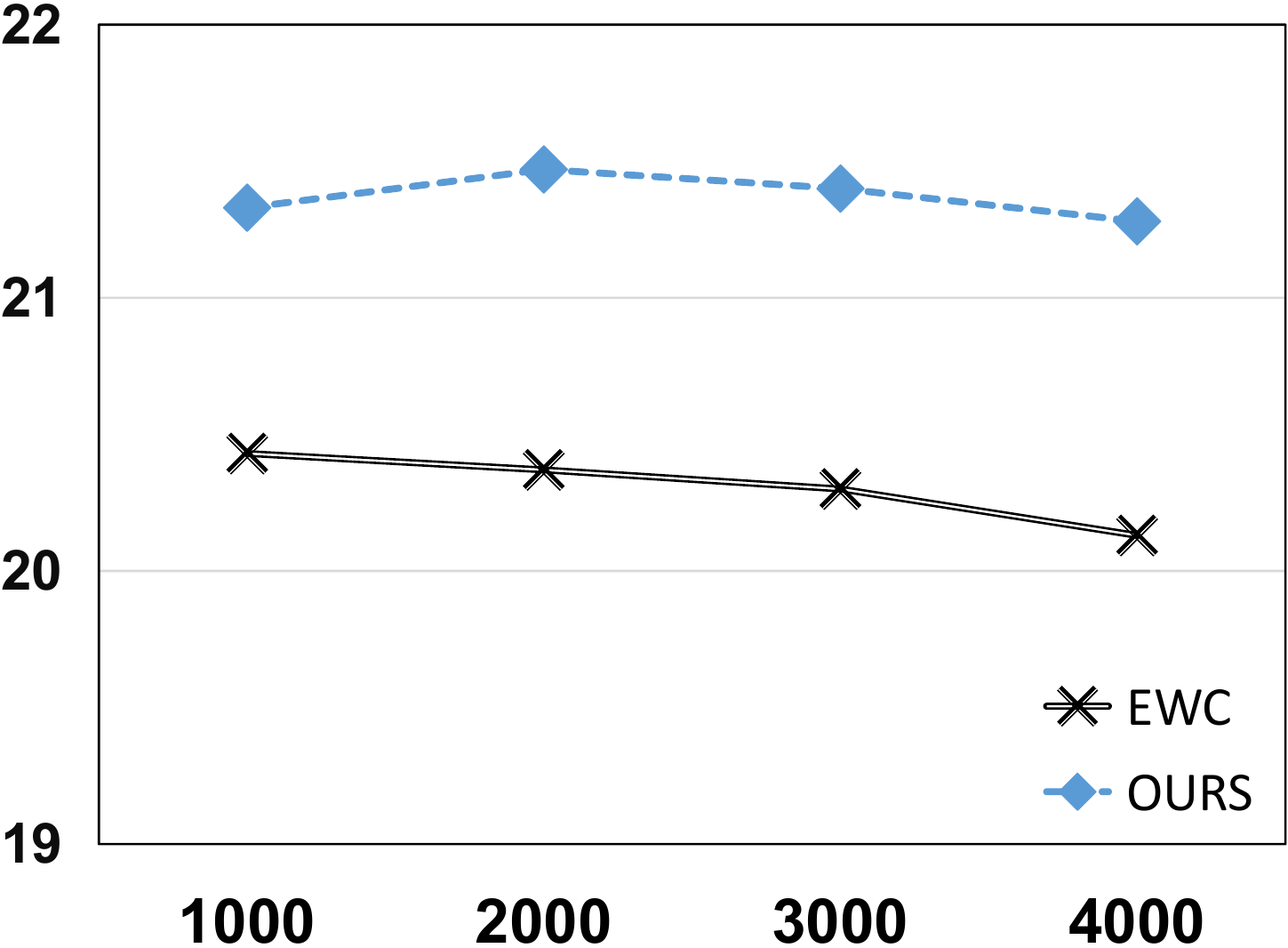}
        \caption{$\lambda$ of CodeBERT.}
        \label{tab:weight-cb}
      \end{subfigure}
      \hfill
      \begin{subfigure}[b]{0.24\textwidth}
        \centering
        \includegraphics[width=1\textwidth]{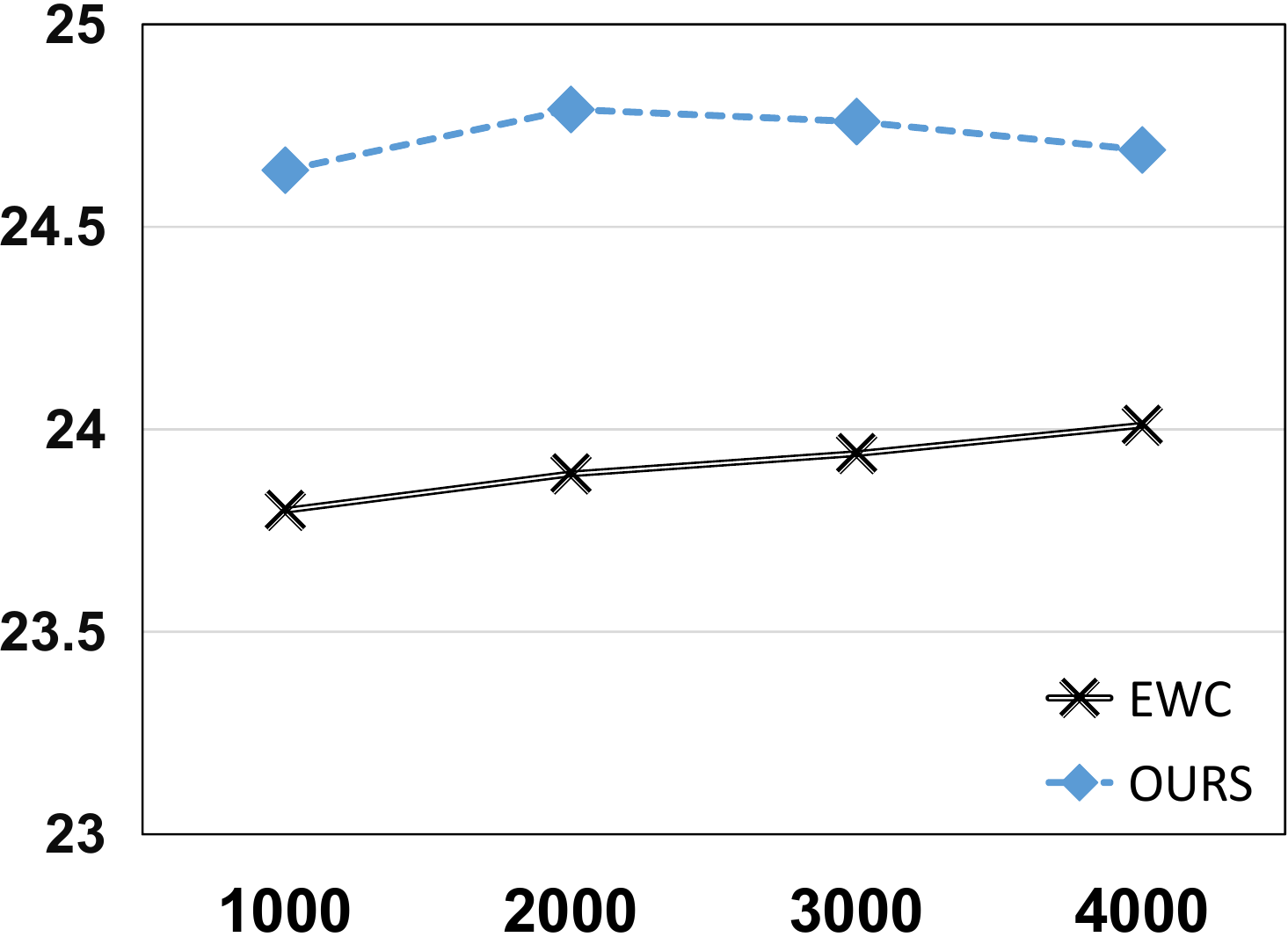}
        \caption{$\lambda$ of CodeT5.}
        \label{tab:weight-ct}
      \end{subfigure}
    \caption{Parameter analysis of exemplars size and $\lambda$ for code summarization. The horizontal axis indicates the value of exemplars size or $\lambda$ while the vertical axis means the BLEU-4 score.}
	\label{fig:parameter1}
\end{figure*}

\begin{figure}
    \centering
    \includegraphics[width=0.49\textwidth]{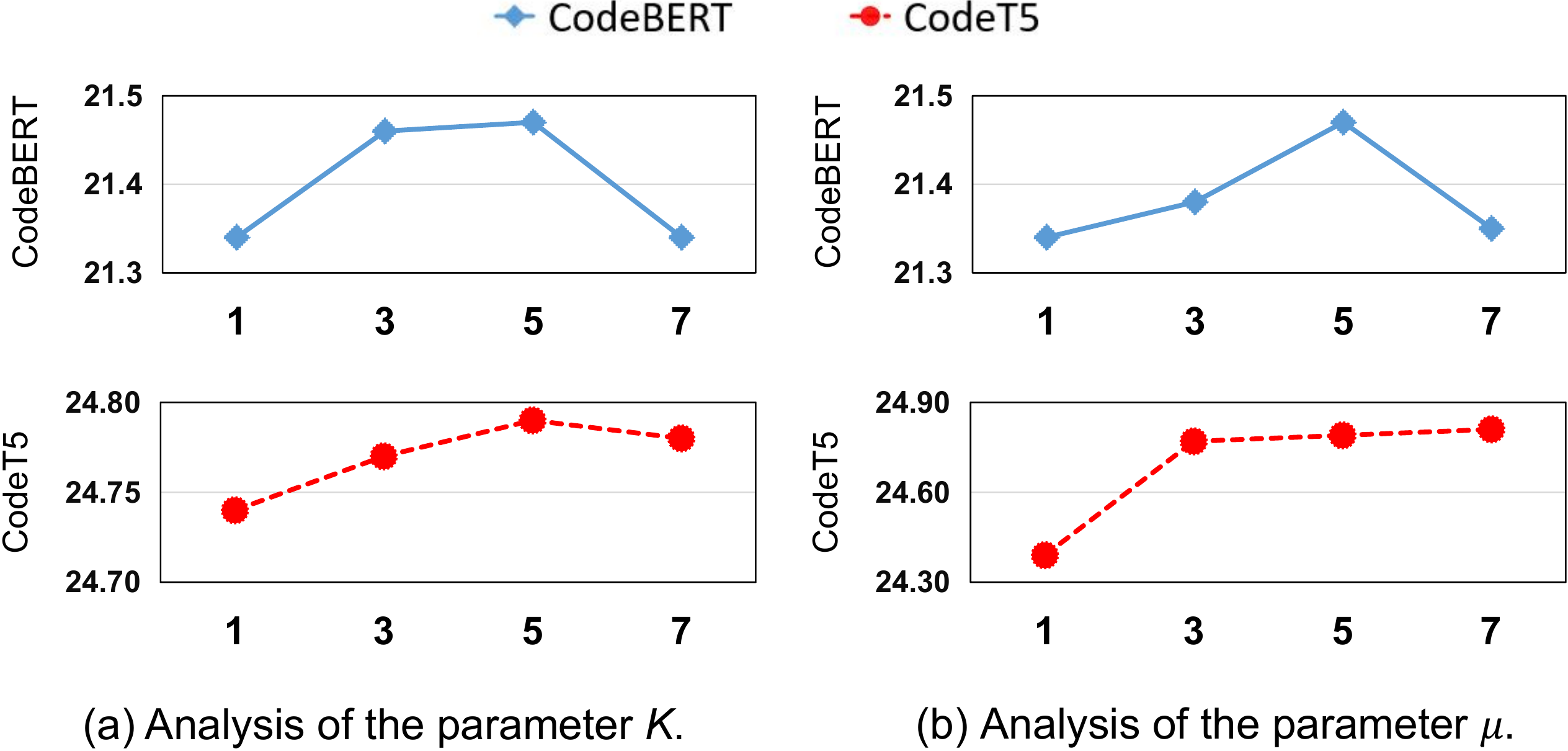}
    \caption{Parameter analysis of (a) $K$ and (b) $\mu$ for code summarization. The horizontal axis indicates the value of parameters while the vertical axis means the BLEU-4 score.}
    \label{fig:parameter2}
\end{figure}

\subsection{RQ3: Ablation Study}\label{sec:RQ3}
We conduct ablation studies to verify the effectiveness of each component in our method, i.e. clustering-based exemplars selection, loss-based exemplars selection, and adaptive regularization. In this experiment, we use both CodeBERT and CodeT5 as our base models and select Java and Python as the evaluation dataset for code summarization. The results on other languages and basic models are presented on our GitHub repository\footnote{{\url{https://github.com/ReliableCoding/REPEAT}}}. 
Table~\ref{tab:ablation} presents the final results averaged by Equ~\ref{equ:ave}. 

\textbf{Clustering-based exemplars selection.} To validate the effectiveness of selecting diverse exemplars, we experiment by removing the clustering process, i.e., setting $K$ to 1. As shown in Table~\ref{tab:ablation}, removing the clustering process dramatically degrades the performance on all tasks. For example, the performance on vulnerability dropped 1.99, 1.37 and 2.16 points 
in terms of F1, Precision, and Recall, respectively, which demonstrates the importance of the diversity of the exemplars. We further show some cases in Section~\ref{sec:case} for illustration.

\textbf{Loss-based exemplars selection.} We conduct this experiment by removing the loss-based exemplar selection in A  lgorithm~\ref{alg:framework}, i.e., selecting exemplars in each cluster randomly. From Table~\ref{tab:ablation}, we can observe that without loss-based exemplars selection, the performance of \tool decreases a lot on all tasks. Specifically, removing this component leads to an obvious decrease in code summarization, with the decrease at 0.64, 0.39, and 0.48 points regarding BLEU-4, METEOR, and ROUGE-L, respectively.  
This indicates the benefits of removing the potential noisy data in exemplars. We also conduct case studies to validate this in Section~\ref{sec:case}. 

\textbf{Adaptive regularization.} We conduct this experiment by removing the cosine similarity term in Equ~\ref{equ:adaptive}. As can be seen in Table~\ref{tab:ablation}, excluding the adaptive regularization leads to a consistent drop in all tasks and metrics. The results demonstrate the effectiveness of adaptively penalizing important parameter change in the continual learning setting.

\subsection{RQ4: Parameter Analysis}\label{sec:RQ4}
In this section, we study the impact of four parameters on results, including two continual learning parameters exemplars size and the weight of EWC $\lambda$, and two hyper-parameters in our method cluster number $K$ and parameter $\mu$ in Algorithm~\ref{alg:framework}. We use the code summarization task and Java dataset for investigation. In each study, we only vary the parameter that needs to be analyzed and keep other parameters unchanged.

\textbf{The exemplars size.}
We conduct experiments to evaluate how EMR and \tool perform under different exemplars size, i.e., 0.1\%, 0.5\%, 1\%, and 2\% of training data. From Figure~\ref{fig:parameter1} (a) and (b), we can observe that \tool outperforms EMR with different exemplar sizes, which demonstrates the effectiveness of \tool. Specifically, \tool improves the BLEU-4 score of EMR by 0.36 and 0.53 points on average for CodeBERT and CodeT5 respectively.

\textbf{The weight of EWC $\lambda$.} To study the impact of $\lambda$ on EWC and \tool, we vary it from 1000 to 4000 and show the results in Figure~\ref{fig:parameter1} (c) and (d). We can find that \tool can outperform EWC with different values of $\lambda$. Since $\lambda$ serves as the degree of penalization on parameter change. Smaller $\lambda$ can not well help the model preserve the previous knowledge, while a larger value of $\lambda$ may harm the learning of new knowledge. For both CodeBERT and CodeT5, \tool can well balance the memory of old knowledge and learning of new knowledge when $\lambda$ is set to 2000. Thus, we select $\lambda$ as 2000 for our method.

\textbf{The parameter $K$.} As shown in Figure~\ref{fig:parameter2} (a), for both CodeBERT and CodeT5, \tool achieves the best performance when $K$ is set to 5. Larger or lower values do not give better results. This indicates that dividing samples into five clusters is more appropriate for \tool. Thus, we set $K$ to 5 in this work.

\textbf{The parameter $\mu$.} Figure~\ref{fig:parameter2} (b) shows the performance variation with the changes of $\mu$. Smaller $\mu$ tends to select the most confident examples in each dataset and larger $\mu$ prefers to select samples more randomly. For CodeBERT, we can find that \tool achieves the best performance when $\mu$ is set to 5. While for the CodeT5, \tool can achieve better performance when $\mu$ is to 5 or 7, with setting to 7 slightly higher than 5. Therefore, we set $\mu$ to 5 to enable \tool to produce relatively better results on both base models.

\section{Discussion}\label{sec:dis}

\subsection{Capability of \tool in other continual learning scenarios}
Experiments in Section~\ref{sec:RQ1} demonstrate the effectiveness of \tool in the project-level continual learning setting. In this section, we further study the performance of \tool in the other two scenarios, i.e., language-level and time-level continual learning. For language-level continual learning, we use the multi-lingual dataset CodeSearchNet for evaluation and train the model in the following order: ``Java $\to$ Python $\to$ Go $\to$ PHP $\to$ Javascript $\to$ Ruby''. Due to the enormous training data, we only use the first part of each language for evaluation. As for the time-level continual learning, we experiment on the Big-Vul~\cite{fan} dataset since only this dataset contains the time information. We split the dataset into five periods: ``Before 2012 $\to$ 2012-2013 $\to$ 2014-2015 $\to$ 2016-2017 $\to$ 2018-2019''. For data in each part, we also randomly split it into training, validation, and test sets with the ratio of 8:1:1.

From Table~\ref{tab:other}, we can find that \tool can also achieve the best performance in these two scenarios. Specifically, for language-level continual learning, \tool improves the best baseline EMR by 0.3 
in terms of BLEU-4 metric. As for time-level continual learning, \tool outperforms other methods by at least 1.65 on the F1 score. The results demonstrate the effectiveness and flexibility of \tool in various continual learning settings.

\begin{table}[t]
    \centering
    \caption{Evaluation of language-level and time-level continual leaning settings with CodeBERT as base model.}
 \scalebox{0.93}{
    \begin{tabular}{c|ccc|ccc}
    \toprule
    \multirow{2}{*}{\textbf{Approach}} & \multicolumn{3}{c|}{Language-level} & \multicolumn{3}{c}{Time-level} \\
    \cmidrule{2-7}
    & BLEU-4 & METEOR & ROUGE-L & F1 & P & R \\
    \midrule
    Upper & 25.46 & 16.54 & 40.17 & 44.79 & 55.95 & 37.50  \\
    \midrule
    FT & 19.35 & 13.23 & 31.87 & 32.50 & 48.59 & 25.07 \\
    \tool & \textbf{22.67} & 14.54 & \textbf{36.30} & \textbf{41.72} & 60.10 & \textbf{32.73}  \\
    \midrule
    EMR & 22.37 & \textbf{14.87} & 36.26 & 40.05 & 53.75 & 32.31  \\
    EWC & 19.57 & 13.29 & 32.39 & 37.35 & \textbf{61.66} & 27.27 \\
    \bottomrule
    \end{tabular}
    \label{tab:other}}
\end{table}

\begin{table}[t]
    \centering
    \caption{Analysis of the training cost of each method for code summarization.}
 \scalebox{1}{
    \begin{tabular}{c|r|r}
    \toprule
    \textbf{Approach} & \multicolumn{1}{c|}{CodeBERT} & \multicolumn{1}{c}{CodeT5} \\
    \midrule
    FT & 5min17s & 6min34s \\
    EMR & 5min35s & 7min01s \\
    EWC & 6min53s & 9min21s \\
    \tool & 7min12s & 9min50s \\
    Upper & 28min20s & 45min40s \\
    \bottomrule
    \end{tabular}
    \label{tab:cost}}
\end{table}

\subsection{Analysis of the training cost of continual learning}\label{sec:cost}
In this section, we investigate the training cost of each method to evaluate whether or not continual learning brings a large training cost. Specifically, we compare the training time per epoch of each method. We use the code summarization task and the fifth part of the Java dataset for investigation. As shown in Table~\ref{tab:cost}, continual learning methods, i.e., EMR, EWC, and \tool, do not increase 
much training time cost comparing
with FT. For example, the time cost of \tool is about 2 and 3 minutes longer than FT on CodeBERT and CodeT5, respectively. We can also observe that Upper dramatically increases the model's training time. Specifically, the training time of Upper is about six times longer than FT on both CodeBERT and CodeT5, which is linear with the number of datasets. This indicates that it is 
hard to use Upper in practice due to the enormous training cost brought by a large number of datasets.

\subsection{Why does \tool work?}\label{sec:case}
\begin{figure}
    \centering
    \includegraphics[width=0.49\textwidth]{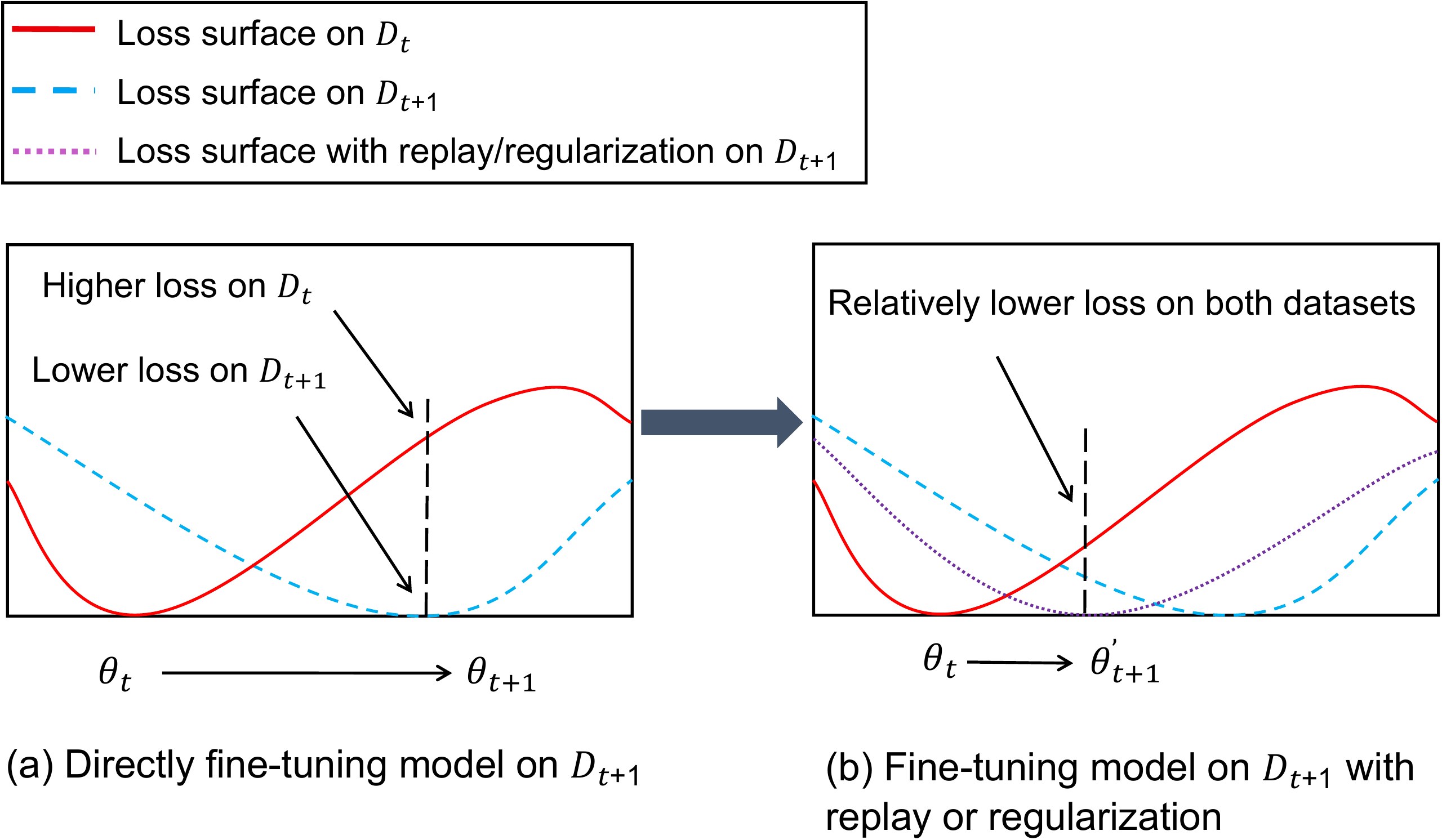}
    \caption{Illustration of how exemplars replay and parameter regularization mitigate the catastrophic forgetting problem.}
    \label{fig:illustration}
\end{figure}

The advantages of our method mainly come from two aspects, exemplars replay and parameter regularization. As illustrated in Figure~\ref{fig:illustration} (a), from the machine learning perspective, the catastrophic forgetting problem is induced by the discrepancy of loss surface in different datasets. For example, when we train the model on the dataset $t$, the model is optimized to converge to $\theta_t$ to obtain the lowest training loss. However, when we further fine-tune the model on dataset $t+1$, the model will be optimized towards $\theta_{t+1}$, which has a lower loss on dataset $t+1$ but a higher loss on dataset $t$. This makes the model forget the knowledge learned before. To mitigate this problem, exemplars replay and parameter regularization work by modifying the loss surface in two ways, i.e., involving previous data and adding a regularization term. In this way, the model will be optimized to $\theta^{'}_{t+1}$, which achieves relatively lower loss on both datasets, as depicted in Figure~\ref{fig:illustration}~(b).

\begin{table}[t]
\newcommand{\tabincell}[2]{\begin{tabular}{@{}#1@{}}#2\end{tabular}}
\centering
\caption{Examples of exemplars selected by \tool w/o clustering and \tool w/o loss.}\label{tab:case_study}
\scalebox{0.93}{
\begin{tabular}{l}
\toprule
 \textbf{Exemplar (1) selected by \tool w/o loss in Go:} \\
 \tabincell{l}{
 \textbf{Code}:\\
 \texttt{\textcolor[RGB]{59,162,59}{\textbf{func}} \textcolor{blue}{NewBalances}(email string, m map[string] string)} \\ \texttt{(*Balances, error)\{} \\ \hspace{2em} \texttt{if email == ""} \\ \hspace{4em}  \texttt{return ...} \\ \texttt{\}} \\
 \textbf{Summarization}: Temporary hack to create an assembly from its id. This is \\ used by SetStatus. We need add a Notifier interface duck typed by Box and \\ Carton?}  \\
\midrule
 \textbf{Example (2) selected by \tool w/o clustering in Go:} \\
 \tabincell{l}{
 \textbf{Code}:\\
 \texttt{\textcolor[RGB]{59,162,59}{\textbf{func}}(in *ConnectionState)\textcolor{blue}{DeepCopy}() *ConnectionState\{} \\ \hspace{2em} \texttt{if in == nil \{return nil\}} \\ \hspace{2em} \texttt{out:=new(ConnectionState)} \\ \hspace{2em} \texttt{in.DeepCopyInto(out)} \\ \hspace{2em} \texttt{return out} \\ \texttt{\}} \\
 \textbf{Summarization}: DeepCopy is an autogenerated deepcopy function \\ copying the receiver creating a new ConnectionState.}  \\
\midrule
 \textbf{Example (3) selected by \tool w/o clustering in Go:} \\
 \tabincell{l}{
 \textbf{Code}:\\
 \texttt{\textcolor[RGB]{59,162,59}{\textbf{func}}(in *ResourceRef)\textcolor{blue}{DeepCopy}() *ResourceRef\{} \\ \hspace{2em} \texttt{if in == nil \{return nil\}} \\ \hspace{2em} \texttt{out:=new(ResourceRef )} \\ \hspace{2em} \texttt{in.DeepCopyInto(out)} \\ \hspace{2em} \texttt{return out} \\ \texttt{\}} \\
 \textbf{Summarization}: DeepCopy is an autogenerated deepcopy function \\ copying the receiver creating a new ResourceRef.}  \\
\bottomrule
\end{tabular}
}
\end{table}

Compared with previous replay-based and regularization-based methods EMR and EWC, \tool mainly benefits from two aspects, i.e., adaptive regularization and representative exemplar selection. The adaptive regularization can flexibly adjust the strength of regularization based on the similarity between different datasets. As for the representative exemplar selection, \tool can select samples with higher quality and better diversity. For the example in Table~\ref{tab:case_study} (1), 
the sample selected by \tool w/o loss is an interrogation that is mainly used for communication, rather than a summarization. This noisy sample does not contribute much to preserving the knowledge of previous datasets. Besides, we can find in Table~\ref{tab:case_study} (2) and (3) that if we exclude the clustering process, the selected samples easily have the same pattern. 
Specifically, 4.8\% of samples selected by \tool w/o clustering have the same pattern, i.e., ``METHOD\_NAME is an autogenerated function ...''. By involving the clustering process, the ratio will drop to 1.7\%. Due to the page limit, we only show part of the cases in this paper and present the full results in our GitHub repository. 

\subsection{Threats to Validity}
We have identified the following major threats to validity:
\begin{enumerate}
\item 
\textit{Base Models.} In this work, we select two widely-used pre-trained models CodeBERT and CodeT5 for evaluation. To comprehensively evaluate the performance of \tool, more pre-trained models such as GraphCodeBERT \cite{DBLP:conf/iclr/GuoRLFT0ZDSFTDC21} and non-pre-trained models like Transformer should also be considered. In our future work, we will verify whether our proposed approach is also effective on other models.

\item 
\textit{Evaluation tasks.} {In this work, we select three popular code intelligence tasks to evaluate \tool , including code summarization, defect detection, and clone detection. Although \tool shows superior performance on these tasks, other important tasks such as code search~\cite{DBLP:conf/sigsoft/CambroneroLKS019,DBLP:journals/nn/GuLGWZXL21} and commit message generation~\cite{DBLP:conf/kbse/LiuXHLXW18,DBLP:journals/tse/LiuGCNL22} are not evaluated in our experiment. In the future, we will validate \tool on more code comprehension tasks.}

\item 
\textit{Evaluation benchmark.} In this work, we construct our evaluation dataset by re-splitting the CodeSearchNet, Big-Vul, and POJ datasets. However, due to the unavailability of time information in most datasets, the projects in each part may not be in chronological order. This may hinder the influence brought by the topic changing in open source platforms like Github. Besides, we only experiment with the datasets divided into five groups. In the future, we will collect a more realistic benchmark for evaluation.

\end{enumerate}


\section{Conclusion}
In this paper, we investigate code intelligence tasks in the continual learning scenario and propose a novel method named \tool to mitigate the catastrophic forgetting problem for code intelligence models. \tool is a method with representative exemplars replay and adaptive parameter regularization. The evaluation on three popular tasks demonstrates the effectiveness of \tool in mitigating the detrimental catastrophic forgetting issue. 

\textbf{Data availability}: We release our source code, experimental data, and detailed experiment results at \textit{{\url{https://github.com/ReliableCoding/REPEAT}}}.


\bibliographystyle{IEEEtran}
\bibliography{sample.bib}

\begin{thebibliography}{10}
\providecommand{\url}[1]{#1}
\csname url@samestyle\endcsname
\providecommand{\newblock}{\relax}
\providecommand{\bibinfo}[2]{#2}
\providecommand{\BIBentrySTDinterwordspacing}{\spaceskip=0pt\relax}
\providecommand{\BIBentryALTinterwordstretchfactor}{4}
\providecommand{\BIBentryALTinterwordspacing}{\spaceskip=\fontdimen2\font plus
\BIBentryALTinterwordstretchfactor\fontdimen3\font minus
  \fontdimen4\font\relax}
\providecommand{\BIBforeignlanguage}[2]{{%
\expandafter\ifx\csname l@#1\endcsname\relax
\typeout{** WARNING: IEEEtran.bst: No hyphenation pattern has been}%
\typeout{** loaded for the language `#1'. Using the pattern for}%
\typeout{** the default language instead.}%
\else
\language=\csname l@#1\endcsname
\fi
#2}}
\providecommand{\BIBdecl}{\relax}
\BIBdecl

\bibitem{DBLP:conf/iwpc/HuLXLJ18}
X.~Hu, G.~Li, X.~Xia, D.~Lo, and Z.~Jin, ``Deep code comment generation,'' in
  \emph{Proceedings of the 26th Conference on Program Comprehension, {ICPC}
  2018, Gothenburg, Sweden, May 27-28, 2018}, F.~Khomh, C.~K. Roy, and
  J.~Siegmund, Eds.\hskip 1em plus 0.5em minus 0.4em\relax {ACM}, 2018, pp.
  200--210.

\bibitem{DBLP:conf/sigsoft/WangYGP0L22}
C.~Wang, Y.~Yang, C.~Gao, Y.~Peng, H.~Zhang, and M.~R. Lyu, ``No more
  fine-tuning? an experimental evaluation of prompt tuning in code
  intelligence,'' in \emph{Proceedings of the 30th {ACM} Joint European
  Software Engineering Conference and Symposium on the Foundations of Software
  Engineering, {ESEC/FSE} 2022, Singapore, Singapore, November 14-18, 2022},
  A.~Roychoudhury, C.~Cadar, and M.~Kim, Eds.\hskip 1em plus 0.5em minus
  0.4em\relax {ACM}, 2022, pp. 382--394.

\bibitem{DBLP:conf/icse/GuZ018}
X.~Gu, H.~Zhang, and S.~Kim, ``Deep code search,'' in \emph{Proceedings of the
  40th International Conference on Software Engineering, {ICSE} 2018,
  Gothenburg, Sweden, May 27 - June 03, 2018}, M.~Chaudron, I.~Crnkovic,
  M.~Chechik, and M.~Harman, Eds.\hskip 1em plus 0.5em minus 0.4em\relax {ACM},
  2018, pp. 933--944.

\bibitem{DBLP:conf/ijcai/ZanCYLKGWCL22}
D.~Zan, B.~Chen, D.~Yang, Z.~Lin, M.~Kim, B.~Guan, Y.~Wang, W.~Chen, and
  J.~Lou, ``{CERT:} continual pre-training on sketches for library-oriented
  code generation,'' in \emph{Proceedings of the Thirty-First International
  Joint Conference on Artificial Intelligence, {IJCAI} 2022, Vienna, Austria,
  23-29 July 2022}, L.~D. Raedt, Ed.\hskip 1em plus 0.5em minus 0.4em\relax
  ijcai.org, 2022, pp. 2369--2375.

\bibitem{DBLP:conf/acl/AhmadCRC20}
W.~U. Ahmad, S.~Chakraborty, B.~Ray, and K.~Chang, ``A transformer-based
  approach for source code summarization,'' in \emph{Proceedings of the 58th
  Annual Meeting of the Association for Computational Linguistics, {ACL}
  2020}.\hskip 1em plus 0.5em minus 0.4em\relax Association for Computational
  Linguistics, 2020, pp. 4998--5007.

\bibitem{DBLP:journals/corr/abs-2002-08653}
W.~Wang, G.~Li, B.~Ma, X.~Xia, and Z.~Jin, ``Detecting code clones with graph
  neural networkand flow-augmented abstract syntax tree,'' \emph{CoRR}, vol.
  abs/2002.08653, 2020.

\bibitem{eva_clone}
H.~Yu, X.~Hu, G.~Li, Y.~Li, Q.~Wang, and T.~Xie, ``Assessing and improving an
  evaluation dataset for detecting semantic code clones via deep learning,''
  \emph{ACM Trans. Softw. Eng. Methodol.}, vol.~31, no.~4, 2022.

\bibitem{DBLP:journals/tosem/ZouZXLJY21}
D.~Zou, Y.~Zhu, S.~Xu, Z.~Li, H.~Jin, and H.~Ye, ``Interpreting deep
  learning-based vulnerability detector predictions based on heuristic
  searching,'' \emph{{ACM} Trans. Softw. Eng. Methodol.}, vol.~30, no.~2, pp.
  23:1--23:31, 2021.

\bibitem{DBLP:conf/nips/ZhouLSD019}
Y.~Zhou, S.~Liu, J.~K. Siow, X.~Du, and Y.~Liu, ``Devign: Effective
  vulnerability identification by learning comprehensive program semantics via
  graph neural networks,'' in \emph{Advances in Neural Information Processing
  Systems 32: Annual Conference on Neural Information Processing Systems 2019,
  NeurIPS 2019}, 2019, pp. 10\,197--10\,207.

\bibitem{DBLP:conf/ndss/LiZXO0WDZ18}
Z.~Li, D.~Zou, S.~Xu, X.~Ou, H.~Jin, S.~Wang, Z.~Deng, and Y.~Zhong,
  ``Vuldeepecker: {A} deep learning-based system for vulnerability detection,''
  in \emph{25th Annual Network and Distributed System Security Symposium,
  {NDSS} 2018}.\hskip 1em plus 0.5em minus 0.4em\relax The Internet Society,
  2018.

\bibitem{timeline}
\url{https://en.wikipedia.org/wiki/Timeline_of_GitHub}.

\bibitem{Oraclejdk}
Oracle, ``Jdk 18 documentation,''
  \url{https://docs.oracle.com/en/java/javase/18/books.html}.

\bibitem{DBLP:conf/nips/BrownMRSKDNSSAA20}
T.~B. Brown, B.~Mann, N.~Ryder, M.~Subbiah, J.~Kaplan, P.~Dhariwal,
  A.~Neelakantan, P.~Shyam, G.~Sastry, A.~Askell, S.~Agarwal,
  A.~Herbert{-}Voss, G.~Krueger, T.~Henighan, R.~Child, A.~Ramesh, D.~M.
  Ziegler, J.~Wu, C.~Winter, C.~Hesse, M.~Chen, E.~Sigler, M.~Litwin, S.~Gray,
  B.~Chess, J.~Clark, C.~Berner, S.~McCandlish, A.~Radford, I.~Sutskever, and
  D.~Amodei, ``Language models are few-shot learners,'' in \emph{Advances in
  Neural Information Processing Systems 33: Annual Conference on Neural
  Information Processing Systems 2020, NeurIPS 2020, December 6-12, 2020,
  virtual}, H.~Larochelle, M.~Ranzato, R.~Hadsell, M.~Balcan, and H.~Lin, Eds.,
  2020.

\bibitem{DBLP:journals/pami/LangeAMPJLST22}
M.~D. Lange, R.~Aljundi, M.~Masana, S.~Parisot, X.~Jia, A.~Leonardis, G.~G.
  Slabaugh, and T.~Tuytelaars, ``A continual learning survey: Defying
  forgetting in classification tasks,'' \emph{{IEEE} Trans. Pattern Anal. Mach.
  Intell.}, vol.~44, no.~7, pp. 3366--3385, 2022.

\bibitem{DBLP:conf/emnlp/FengGTDFGS0LJZ20}
Z.~Feng, D.~Guo, D.~Tang, N.~Duan, X.~Feng, M.~Gong, L.~Shou, B.~Qin, T.~Liu,
  D.~Jiang, and M.~Zhou, ``Codebert: {A} pre-trained model for programming and
  natural languages,'' in \emph{Findings of the Association for Computational
  Linguistics: {EMNLP} 2020}, ser. Findings of {ACL}, vol. {EMNLP} 2020.\hskip
  1em plus 0.5em minus 0.4em\relax Association for Computational Linguistics,
  2020, pp. 1536--1547.

\bibitem{DBLP:conf/emnlp/0034WJH21}
Y.~Wang, W.~Wang, S.~R. Joty, and S.~C.~H. Hoi, ``Codet5: Identifier-aware
  unified pre-trained encoder-decoder models for code understanding and
  generation,'' in \emph{Proceedings of the 2021 Conference on Empirical
  Methods in Natural Language Processing, {EMNLP} 2021}.\hskip 1em plus 0.5em
  minus 0.4em\relax Association for Computational Linguistics, 2021, pp.
  8696--8708.

\bibitem{DBLP:journals/pami/LiH18a}
Z.~Li and D.~Hoiem, ``Learning without forgetting,'' \emph{{IEEE} Trans.
  Pattern Anal. Mach. Intell.}, vol.~40, no.~12, pp. 2935--2947, 2018.

\bibitem{chen2018lifelong}
Z.~Chen and B.~Liu, ``Lifelong machine learning,'' \emph{Synthesis Lectures on
  Artificial Intelligence and Machine Learning}, vol.~12, no.~3, pp. 1--207,
  2018.

\bibitem{DBLP:conf/naacl/WangXYGCW19}
H.~Wang, W.~Xiong, M.~Yu, X.~Guo, S.~Chang, and W.~Y. Wang, ``Sentence
  embedding alignment for lifelong relation extraction,'' in \emph{Proceedings
  of the 2019 Conference of the North American Chapter of the Association for
  Computational Linguistics: Human Language Technologies, {NAACL-HLT} 2019,
  Minneapolis, MN, USA, June 2-7, 2019, Volume 1 (Long and Short Papers)},
  J.~Burstein, C.~Doran, and T.~Solorio, Eds.\hskip 1em plus 0.5em minus
  0.4em\relax Association for Computational Linguistics, 2019, pp. 796--806.

\bibitem{DBLP:conf/acl/HanDGLLLSZ20}
X.~Han, Y.~Dai, T.~Gao, Y.~Lin, Z.~Liu, P.~Li, M.~Sun, and J.~Zhou, ``Continual
  relation learning via episodic memory activation and reconsolidation,'' in
  \emph{Proceedings of the 58th Annual Meeting of the Association for
  Computational Linguistics, {ACL} 2020, Online, July 5-10, 2020}, D.~Jurafsky,
  J.~Chai, N.~Schluter, and J.~R. Tetreault, Eds.\hskip 1em plus 0.5em minus
  0.4em\relax Association for Computational Linguistics, 2020, pp. 6429--6440.

\bibitem{DBLP:conf/emnlp/MiCZHF20}
F.~Mi, L.~Chen, M.~Zhao, M.~Huang, and B.~Faltings, ``Continual learning for
  natural language generation in task-oriented dialog systems,'' in
  \emph{Findings of the Association for Computational Linguistics: {EMNLP}
  2020, Online Event, 16-20 November 2020}, ser. Findings of {ACL}, T.~Cohn,
  Y.~He, and Y.~Liu, Eds., vol. {EMNLP} 2020.\hskip 1em plus 0.5em minus
  0.4em\relax Association for Computational Linguistics, 2020, pp. 3461--3474.

\bibitem{DBLP:conf/www/YuanYHCWC22}
W.~Yuan, H.~Yin, T.~He, T.~Chen, Q.~Wang, and L.~Cui, ``Unified question
  generation with continual lifelong learning,'' in \emph{{WWW} '22: The {ACM}
  Web Conference 2022, Virtual Event, Lyon, France, April 25 - 29, 2022},
  F.~Laforest, R.~Troncy, E.~Simperl, D.~Agarwal, A.~Gionis, I.~Herman, and
  L.~M{\'{e}}dini, Eds.\hskip 1em plus 0.5em minus 0.4em\relax {ACM}, 2022, pp.
  871--881.

\bibitem{kirkpatrick2017overcoming}
J.~Kirkpatrick, R.~Pascanu, N.~Rabinowitz, J.~Veness, G.~Desjardins, A.~A.
  Rusu, K.~Milan, J.~Quan, T.~Ramalho, A.~Grabska-Barwinska \emph{et~al.},
  ``Overcoming catastrophic forgetting in neural networks,'' \emph{Proceedings
  of the national academy of sciences}, vol. 114, no.~13, pp. 3521--3526, 2017.

\bibitem{DBLP:conf/oopsla/Allamanis19}
M.~Allamanis, ``The adverse effects of code duplication in machine learning
  models of code,'' in \emph{Proceedings of the 2019 {ACM} {SIGPLAN}
  International Symposium on New Ideas, New Paradigms, and Reflections on
  Programming and Software, Onward! 2019, Athens, Greece, October 23-24, 2019},
  H.~Masuhara and T.~Petricek, Eds.\hskip 1em plus 0.5em minus 0.4em\relax
  {ACM}, 2019, pp. 143--153.

\bibitem{DBLP:conf/icse/SunLL0022}
Z.~Sun, L.~Li, Y.~Liu, X.~Du, and L.~Li, ``On the importance of building
  high-quality training datasets for neural code search,'' in \emph{44th
  {IEEE/ACM} 44th International Conference on Software Engineering, {ICSE}
  2022, Pittsburgh, PA, USA, May 25-27, 2022}.\hskip 1em plus 0.5em minus
  0.4em\relax {ACM}, 2022, pp. 1609--1620.

\bibitem{DBLP:journals/corr/abs-2207-05579}
L.~Shi, F.~Mu, X.~Chen, S.~Wang, J.~Wang, Y.~Yang, G.~Li, X.~Xia, and Q.~Wang,
  ``Are we building on the rock? on the importance of data preprocessing for
  code summarization,'' \emph{CoRR}, vol. abs/2207.05579, 2022.

\bibitem{DBLP:journals/tse/KamiyaKI02}
T.~Kamiya, S.~Kusumoto, and K.~Inoue, ``Ccfinder: {A} multilinguistic
  token-based code clone detection system for large scale source code,''
  \emph{{IEEE} Trans. Software Eng.}, vol.~28, no.~7, pp. 654--670, 2002.

\bibitem{DBLP:conf/sigsoft/KimSN05}
M.~Kim, V.~Sazawal, D.~Notkin, and G.~C. Murphy, ``An empirical study of code
  clone genealogies,'' in \emph{Proceedings of the 10th European Software
  Engineering Conference held jointly with 13th {ACM} {SIGSOFT} International
  Symposium on Foundations of Software Engineering, 2005, Lisbon, Portugal,
  September 5-9, 2005}, M.~Wermelinger and H.~C. Gall, Eds.\hskip 1em plus
  0.5em minus 0.4em\relax {ACM}, 2005, pp. 187--196.

\bibitem{DBLP:conf/acl/IyerKCZ16}
S.~Iyer, I.~Konstas, A.~Cheung, and L.~Zettlemoyer, ``Summarizing source code
  using a neural attention model,'' in \emph{Proceedings of the 54th Annual
  Meeting of the Association for Computational Linguistics, {ACL} 2016}.\hskip
  1em plus 0.5em minus 0.4em\relax The Association for Computer Linguistics,
  2016.

\bibitem{DBLP:conf/acl/WuZZ21}
H.~Wu, H.~Zhao, and M.~Zhang, ``Code summarization with structure-induced
  transformer,'' in \emph{Findings of the Association for Computational
  Linguistics: {ACL/IJCNLP} 2021, Online Event, August 1-6, 2021}, ser.
  Findings of {ACL}, C.~Zong, F.~Xia, W.~Li, and R.~Navigli, Eds., vol.
  {ACL/IJCNLP} 2021.\hskip 1em plus 0.5em minus 0.4em\relax Association for
  Computational Linguistics, 2021, pp. 1078--1090.

\bibitem{DBLP:conf/icse/TangSLGHZ022}
Z.~Tang, X.~Shen, C.~Li, J.~Ge, L.~Huang, Z.~Zhu, and B.~Luo, ``Ast-trans: Code
  summarization with efficient tree-structured attention,'' in \emph{44th
  {IEEE/ACM} 44th International Conference on Software Engineering, {ICSE}
  2022, Pittsburgh, PA, USA, May 25-27, 2022}.\hskip 1em plus 0.5em minus
  0.4em\relax {IEEE}, 2022, pp. 150--162.

\bibitem{10.1145/3522674}
S.~Gao, C.~Gao, Y.~He, J.~Zeng, L.~Y. Nie, X.~Xia, and M.~R. Lyu, ``Code
  structure guided transformer for source code summarization,'' \emph{ACM
  Trans. Softw. Eng. Methodol.}, 2022.

\bibitem{DBLP:journals/corr/abs-2102-04664}
S.~Lu, D.~Guo, S.~Ren, J.~Huang, A.~Svyatkovskiy, A.~Blanco, C.~B. Clement,
  D.~Drain, D.~Jiang, D.~Tang, G.~Li, L.~Zhou, L.~Shou, L.~Zhou, M.~Tufano,
  M.~Gong, M.~Zhou, N.~Duan, N.~Sundaresan, S.~K. Deng, S.~Fu, and S.~Liu,
  ``Codexglue: {A} machine learning benchmark dataset for code understanding
  and generation,'' \emph{CoRR}, vol. abs/2102.04664, 2021.

\bibitem{DBLP:journals/corr/abs-2212-14274}
X.~Wen, C.~Gao, J.~Ye, Z.~Tian, Y.~Jia, and X.~Wang, ``Meta-path based
  attentional graph learning model for vulnerability detection,'' \emph{CoRR},
  vol. abs/2212.14274, 2022.

\bibitem{DBLP:conf/ijcai/WeiL17}
H.~Wei and M.~Li, ``Supervised deep features for software functional clone
  detection by exploiting lexical and syntactical information in source code,''
  in \emph{Proceedings of the Twenty-Sixth International Joint Conference on
  Artificial Intelligence, {IJCAI} 2017, Melbourne, Australia, August 19-25,
  2017}, C.~Sierra, Ed.\hskip 1em plus 0.5em minus 0.4em\relax ijcai.org, 2017,
  pp. 3034--3040.

\bibitem{DBLP:conf/iwpc/YuLCLXW19}
H.~Yu, W.~Lam, L.~Chen, G.~Li, T.~Xie, and Q.~Wang, ``Neural detection of
  semantic code clones via tree-based convolution,'' in \emph{Proceedings of
  the 27th International Conference on Program Comprehension, {ICPC} 2019,
  Montreal, QC, Canada, May 25-31, 2019}, Y.~Gu{\'{e}}h{\'{e}}neuc, F.~Khomh,
  and F.~Sarro, Eds.\hskip 1em plus 0.5em minus 0.4em\relax {IEEE} / {ACM},
  2019, pp. 70--80.

\bibitem{DBLP:conf/iclr/GuoRLFT0ZDSFTDC21}
D.~Guo, S.~Ren, S.~Lu, Z.~Feng, D.~Tang, S.~Liu, L.~Zhou, N.~Duan,
  A.~Svyatkovskiy, S.~Fu, M.~Tufano, S.~K. Deng, C.~B. Clement, D.~Drain,
  N.~Sundaresan, J.~Yin, D.~Jiang, and M.~Zhou, ``Graphcodebert: Pre-training
  code representations with data flow,'' in \emph{9th International Conference
  on Learning Representations, {ICLR} 2021}.\hskip 1em plus 0.5em minus
  0.4em\relax OpenReview.net, 2021.

\bibitem{mccloskey1989catastrophic}
M.~McCloskey and N.~J. Cohen, ``Catastrophic interference in connectionist
  networks: The sequential learning problem,'' in \emph{Psychology of learning
  and motivation}.\hskip 1em plus 0.5em minus 0.4em\relax Elsevier, 1989,
  vol.~24, pp. 109--165.

\bibitem{DBLP:conf/nips/Lopez-PazR17}
D.~Lopez{-}Paz and M.~Ranzato, ``Gradient episodic memory for continual
  learning,'' in \emph{Advances in Neural Information Processing Systems 30:
  Annual Conference on Neural Information Processing Systems 2017, December
  4-9, 2017, Long Beach, CA, {USA}}, I.~Guyon, U.~von Luxburg, S.~Bengio, H.~M.
  Wallach, R.~Fergus, S.~V.~N. Vishwanathan, and R.~Garnett, Eds., 2017, pp.
  6467--6476.

\bibitem{DBLP:conf/iclr/SunHL20}
F.~Sun, C.~Ho, and H.~Lee, ``{LAMOL:} language modeling for lifelong language
  learning,'' in \emph{8th International Conference on Learning
  Representations, {ICLR} 2020, Addis Ababa, Ethiopia, April 26-30,
  2020}.\hskip 1em plus 0.5em minus 0.4em\relax OpenReview.net, 2020.

\bibitem{DBLP:conf/cvpr/AljundiCT17}
R.~Aljundi, P.~Chakravarty, and T.~Tuytelaars, ``Expert gate: Lifelong learning
  with a network of experts,'' in \emph{2017 {IEEE} Conference on Computer
  Vision and Pattern Recognition, {CVPR} 2017, Honolulu, HI, USA, July 21-26,
  2017}.\hskip 1em plus 0.5em minus 0.4em\relax {IEEE} Computer Society, 2017,
  pp. 7120--7129.

\bibitem{DBLP:conf/cvpr/MallyaL18}
A.~Mallya and S.~Lazebnik, ``Packnet: Adding multiple tasks to a single network
  by iterative pruning,'' in \emph{2018 {IEEE} Conference on Computer Vision
  and Pattern Recognition, {CVPR} 2018, Salt Lake City, UT, USA, June 18-22,
  2018}.\hskip 1em plus 0.5em minus 0.4em\relax Computer Vision Foundation /
  {IEEE} Computer Society, 2018, pp. 7765--7773.

\bibitem{DBLP:conf/iclr/RamalhoG19}
T.~Ramalho and M.~Garnelo, ``Adaptive posterior learning: few-shot learning
  with a surprise-based memory module,'' in \emph{7th International Conference
  on Learning Representations, {ICLR} 2019, New Orleans, LA, USA, May 6-9,
  2019}.\hskip 1em plus 0.5em minus 0.4em\relax OpenReview.net, 2019.

\bibitem{DBLP:conf/iclr/KemkerK18}
R.~Kemker and C.~Kanan, ``Fearnet: Brain-inspired model for incremental
  learning,'' in \emph{6th International Conference on Learning
  Representations, {ICLR} 2018, Vancouver, BC, Canada, April 30 - May 3, 2018,
  Conference Track Proceedings}.\hskip 1em plus 0.5em minus 0.4em\relax
  OpenReview.net, 2018.

\bibitem{DBLP:conf/emnlp/LiQH21}
Z.~Li, L.~Qu, and G.~Haffari, ``Total recall: a customized continual learning
  method for neural semantic parsers,'' in \emph{Proceedings of the 2021
  Conference on Empirical Methods in Natural Language Processing, {EMNLP} 2021,
  Virtual Event / Punta Cana, Dominican Republic, 7-11 November, 2021},
  M.~Moens, X.~Huang, L.~Specia, and S.~W. Yih, Eds.\hskip 1em plus 0.5em minus
  0.4em\relax Association for Computational Linguistics, 2021, pp. 3816--3831.

\bibitem{DBLP:conf/issta/YuanZHFHHY22}
W.~Yuan, Q.~Zhang, T.~He, C.~Fang, N.~Q.~V. Hung, X.~Hao, and H.~Yin,
  ``{CIRCLE:} continual repair across programming languages,'' in \emph{{ISSTA}
  '22: 31st {ACM} {SIGSOFT} International Symposium on Software Testing and
  Analysis, Virtual Event, South Korea, July 18 - 22, 2022}, S.~Ryu and
  Y.~Smaragdakis, Eds.\hskip 1em plus 0.5em minus 0.4em\relax {ACM}, 2022, pp.
  678--690.

\bibitem{DBLP:conf/naacl/LeClairM19}
A.~LeClair and C.~McMillan, ``Recommendations for datasets for source code
  summarization,'' in \emph{Proceedings of the 2019 Conference of the North
  American Chapter of the Association for Computational Linguistics: Human
  Language Technologies, {NAACL-HLT} 2019, Minneapolis, MN, USA, June 2-7,
  2019, Volume 1 (Long and Short Papers)}, J.~Burstein, C.~Doran, and
  T.~Solorio, Eds.\hskip 1em plus 0.5em minus 0.4em\relax Association for
  Computational Linguistics, 2019, pp. 3931--3937.

\bibitem{DBLP:conf/kbse/WeiLLXJ20}
B.~Wei, Y.~Li, G.~Li, X.~Xia, and Z.~Jin, ``Retrieve and refine: Exemplar-based
  neural comment generation,'' in \emph{35th {IEEE/ACM} International
  Conference on Automated Software Engineering, {ASE} 2020}.\hskip 1em plus
  0.5em minus 0.4em\relax {IEEE}, 2020, pp. 349--360.

\bibitem{ramos2003using}
J.~Ramos \emph{et~al.}, ``Using tf-idf to determine word relevance in document
  queries,'' in \emph{Proceedings of the first instructional conference on
  machine learning}, vol. 242, no.~1.\hskip 1em plus 0.5em minus 0.4em\relax
  Citeseer, 2003, pp. 29--48.

\bibitem{DBLP:conf/nips/HanYYNXHTS18}
B.~Han, Q.~Yao, X.~Yu, G.~Niu, M.~Xu, W.~Hu, I.~W. Tsang, and M.~Sugiyama,
  ``Co-teaching: Robust training of deep neural networks with extremely noisy
  labels,'' in \emph{Advances in Neural Information Processing Systems 31:
  Annual Conference on Neural Information Processing Systems 2018, NeurIPS
  2018, December 3-8, 2018, Montr{\'{e}}al, Canada}, 2018, pp. 8536--8546.

\bibitem{DBLP:conf/iccv/HuangQJZ19}
J.~Huang, L.~Qu, R.~Jia, and B.~Zhao, ``O2u-net: {A} simple noisy label
  detection approach for deep neural networks,'' in \emph{2019 {IEEE/CVF}
  International Conference on Computer Vision, {ICCV} 2019, Seoul, Korea
  (South), October 27 - November 2, 2019}.\hskip 1em plus 0.5em minus
  0.4em\relax {IEEE}, 2019, pp. 3325--3333.

\bibitem{DBLP:journals/corr/abs-1909-09436}
H.~Husain, H.~Wu, T.~Gazit, M.~Allamanis, and M.~Brockschmidt, ``Codesearchnet
  challenge: Evaluating the state of semantic code search,'' \emph{CoRR}, vol.
  abs/1909.09436, 2019.

\bibitem{fan}
J.~Fan, Y.~Li, S.~Wang, and T.~N. Nguyen, ``A {C/C++} code vulnerability
  dataset with code changes and {CVE} summaries,'' in \emph{{MSR} '20: 17th
  International Conference on Mining Software Repositories, Seoul, Republic of
  Korea, 29-30 June, 2020}.\hskip 1em plus 0.5em minus 0.4em\relax {ACM}, 2020,
  pp. 508--512.

\bibitem{DBLP:conf/aaai/MouLZWJ16}
L.~Mou, G.~Li, L.~Zhang, T.~Wang, and Z.~Jin, ``Convolutional neural networks
  over tree structures for programming language processing,'' in
  \emph{Proceedings of the Thirtieth {AAAI} Conference on Artificial
  Intelligence}.\hskip 1em plus 0.5em minus 0.4em\relax {AAAI} Press, 2016, pp.
  1287--1293.

\bibitem{DBLP:conf/iclr/LiuCXS021}
S.~Liu, Y.~Chen, X.~Xie, J.~K. Siow, and Y.~Liu, ``Retrieval-augmented
  generation for code summarization via hybrid {GNN},'' in \emph{9th
  International Conference on Learning Representations, {ICLR} 2021}.\hskip 1em
  plus 0.5em minus 0.4em\relax OpenReview.net, 2021.

\bibitem{DBLP:conf/icse/ZhangW00020}
J.~Zhang, X.~Wang, H.~Zhang, H.~Sun, and X.~Liu, ``Retrieval-based neural
  source code summarization,'' in \emph{{ICSE} '20: 42nd International
  Conference on Software Engineering, Seoul, South Korea, 27 June - 19 July,
  2020}, G.~Rothermel and D.~Bae, Eds.\hskip 1em plus 0.5em minus 0.4em\relax
  {ACM}, 2020, pp. 1385--1397.

\bibitem{DBLP:conf/acl/PapineniRWZ02}
K.~Papineni, S.~Roukos, T.~Ward, and W.~Zhu, ``Bleu: a method for automatic
  evaluation of machine translation,'' in \emph{Proceedings of the 40th Annual
  Meeting of the Association for Computational Linguistics, July 6-12, 2002,
  Philadelphia, PA, {USA}}.\hskip 1em plus 0.5em minus 0.4em\relax {ACL}, 2002,
  pp. 311--318.

\bibitem{lin-2004-rouge}
C.-Y. Lin, ``{ROUGE}: A package for automatic evaluation of summaries,'' in
  \emph{Text Summarization Branches Out}.\hskip 1em plus 0.5em minus
  0.4em\relax Barcelona, Spain: Association for Computational Linguistics, Jul.
  2004, pp. 74--81.

\bibitem{DBLP:conf/acl/BanerjeeL05}
S.~Banerjee and A.~Lavie, ``{METEOR:} an automatic metric for {MT} evaluation
  with improved correlation with human judgments,'' in \emph{Proceedings of the
  Workshop on Intrinsic and Extrinsic Evaluation Measures for Machine
  Translation and/or Summarization@ACL 2005, Ann Arbor, Michigan, USA, June 29,
  2005}, J.~Goldstein, A.~Lavie, C.~Lin, and C.~R. Voss, Eds.\hskip 1em plus
  0.5em minus 0.4em\relax Association for Computational Linguistics, 2005, pp.
  65--72.

\bibitem{DBLP:conf/sigsoft/CambroneroLKS019}
J.~Cambronero, H.~Li, S.~Kim, K.~Sen, and S.~Chandra, ``When deep learning met
  code search,'' in \emph{Proceedings of the {ACM} Joint Meeting on European
  Software Engineering Conference and Symposium on the Foundations of Software
  Engineering, {ESEC/SIGSOFT} {FSE} 2019, Tallinn, Estonia, August 26-30,
  2019}, M.~Dumas, D.~Pfahl, S.~Apel, and A.~Russo, Eds.\hskip 1em plus 0.5em
  minus 0.4em\relax {ACM}, 2019, pp. 964--974.

\bibitem{DBLP:journals/nn/GuLGWZXL21}
W.~Gu, Z.~Li, C.~Gao, C.~Wang, H.~Zhang, Z.~Xu, and M.~R. Lyu, ``Cradle: Deep
  code retrieval based on semantic dependency learning,'' \emph{Neural
  Networks}, vol. 141, pp. 385--394, 2021.

\bibitem{DBLP:conf/kbse/LiuXHLXW18}
Z.~Liu, X.~Xia, A.~E. Hassan, D.~Lo, Z.~Xing, and X.~Wang,
  ``Neural-machine-translation-based commit message generation: how far are
  we?'' in \emph{Proceedings of the 33rd {ACM/IEEE} International Conference on
  Automated Software Engineering, {ASE} 2018, Montpellier, France, September
  3-7, 2018}, M.~Huchard, C.~K{\"{a}}stner, and G.~Fraser, Eds.\hskip 1em plus
  0.5em minus 0.4em\relax {ACM}, 2018, pp. 373--384.

\bibitem{DBLP:journals/tse/LiuGCNL22}
S.~Liu, C.~Gao, S.~Chen, L.~Y. Nie, and Y.~Liu, ``{ATOM:} commit message
  generation based on abstract syntax tree and hybrid ranking,'' \emph{{IEEE}
  Trans. Software Eng.}, vol.~48, no.~5, pp. 1800--1817, 2022.

\end{thebibliography}

\end{document}